\documentclass[reprint,10pt,superscriptaddress,twocolumn,aps,prl]{revtex4-2}
\usepackage{amsmath}
\usepackage{amssymb}
\usepackage{graphicx}
\usepackage[colorlinks=true,citecolor=blue,linktocpage=true,linkcolor=blue,filecolor=blue,urlcolor=magenta]{hyperref}
\usepackage{physics}
\usepackage{bbm}
\usepackage{color}
\usepackage{dcolumn}
\usepackage{color}

\newcommand{\beginsupplement}{%
            \setcounter{table}{0}
            \renewcommand{\thetable}{S\arabic{table}}%
            \setcounter{equation}{0}
            \renewcommand{\theequation}{S\arabic{equation}}%
            \setcounter{figure}{0}
            \renewcommand{\thefigure}{S\arabic{figure}}%
}

\makeatletter
\def\maketitle{
\@author@finish
\title@column\titleblock@produce
\suppressfloats[t]}
\makeatother

\begin{document}

\preprint{APS/123-QED}

\title{Semi-Analytical Engineering of Strongly Driven Nonlinear Systems\\ Beyond Floquet and Perturbation Theory}

\author{Kento~Taniguchi}
 \email{kento-taniguchi531@berkeley.edu}
 \altaffiliation[Present address: ]{Department of Physics, University of California, Berkeley, Berkeley, CA 94720, USA}
 \affiliation{Komaba Institute for Science (KIS), The University of Tokyo, Meguro-ku, Tokyo, 153-8902, Japan}%

\author{Atsushi~Noguchi}
\affiliation{Komaba Institute for Science (KIS), The University of Tokyo, Meguro-ku, Tokyo, 153-8902, Japan}
\affiliation{RIKEN Center for Quantum Computing (RQC), RIKEN, Wako-shi, Saitama 351-0198, Japan}
\affiliation{Inamori Research Institute for Science (InaRIS), Kyoto-shi, Kyoto, 600-8411, Japan}%

\author{Takashi~Oka}
\affiliation{The Institute of Solid State Physics, The University of Tokyo, Kashiwa, Chiba 277-8581, Japan}%

\date{\today}

\begin{abstract}
Strongly driven nonlinear systems are frequently encountered in physics, yet their accurate control is generally challenging due to the intricate dynamics. In this work, we present a non-perturbative, semi-analytical framework for tailoring such systems. The key idea is heuristically extending the Floquet theory to nonlinear differential equations using the Harmonic Balance method. Additionally, we establish a novel constrained optimization technique inspired by the Lagrange multiplier method. This approach enables accurate engineering of effective potentials across a broader parameter space, surpassing the limitations of perturbative methods. Our method offers practical implementations in diverse experimental platforms, facilitating nonclassical state generation, versatile bosonic quantum simulations, and solving complex optimization problems across quantum and classical applications.

\end{abstract}

\maketitle


\textit{Introduction}---Periodically driven nonlinear systems exhibit rich and non-trivial phenomena, offering the potential for breakthroughs in quantum technologies. These systems, exemplified by high harmonic generation \cite{Winterfeldt2008}, time crystals \cite{Zaletel2023}, Floquet topological phases \cite{Eckardt2017}, and chaos-assisted dynamical tunneling \cite{Hensinger2001}, provide a fertile platform for exploring new frontiers. Recent developments in quantum technologies have demonstrated their ability in Hamiltonian engineering to produce nonclassical states across various experiments, including nuclear magnetic resonance \cite{Choi2020}, superconducting circuits \cite{nguyen2024}, Paul traps \cite{Kiefer2019}, and driven optical lattices \cite{Yin2022}. 

A central and unsolved challenge in this field is solving inverse problems \cite{Guo2024}: how to optimally design a time-dependent control Hamiltonian that realizes a specified, time-independent effective Hamiltonian. This issue lies at the heart of Floquet engineering, which uses time-periodic modulation to control a system leveraging Floquet theory and perturbative techniques \cite{Oka2019}. However, conventional Floquet engineering has been limited to systems governed by linear differential equations, such as the Schr\"{o}dinger equation, due to the linearity requirements of Floquet theory \cite{Goldman2014}. In contrast, many practical systems --- including driven anharmonic oscillators, Bose--Einstein condensates, and micromagnets --- are described by nonlinear differential equations, such as the nonlinear Mathieu equation \cite{kovacic2018}, the Gross--Pitaevskii equation \cite{Stamper2013}, and the Landau--Lifshitz--Gilbert equation \cite{Gilbert2004}. Extending Floquet engineering to encompass these nonlinear systems is thus of great importance.  

Recent progress has extended Floquet engineering to the nonlinear differential equations by utilizing the linearity of the corresponding Liouville's equation in a perturbative framework \cite{Martin2015, higashikawa2018, Mori2018, Mori2022}. However, solving general inverse problems with accuracy and efficiency remains a theoretical challenge \cite{Mori2018}. The difficulty lies in the growing complexity of higher-order corrections, which constrain the precision of perturbative approaches and limit their applicability to weak and fast drive conditions \cite{Sun2020, Penin2024}. Alternative methods, such as reinforcement learning and optimal control theory, have achieved success in applications like band structure engineering and nonclassical state generation \cite{Werschnik2007, Bukov2018, Castro2022, Matsos2024, VGMatsos2024}, but they are computationally intensive and often obscure the underlying mechanisms. Meanwhile, rapidly developing quantum technologies keep breaking the records of high-fidelity quantum operations in diverse platforms \cite{evered2023, manetsch2024, Li2024, loschnauer2024}, increasing the demand for more precise Hamiltonian implementations. Therefore, we need a novel framework of Floquet engineering that combines accuracy, computational efficiency, and broader applicability.

The Multidimensional Alternating Frequency/Time Domain Harmonic Balance (MAFT-HB) method is a semi-analytical method used for nonlinear studies \cite{Junge2021}. In this letter, we leverage the MAFT-HB method for heuristically extending the Floquet theory to nonlinear differential equations. Based on this, we establish a novel constrained optimization technique, inspired by the Lagrange multiplier method \cite{Rockafellar1993}, to address inverse problems in strongly driven nonlinear systems. Unlike previous research, our method fundamentally eliminates the linearity assumption of Floquet theory, enabling its direct applications to nonlinear differential equations. Moreover, the synergy of non-perturbative, semi-analytical frameworks realizes efficient and accurate resolution of inverse problems, as exemplified by the inverse scattering transform in Soliton research \cite{ablowitz1981}, providing deep insight into the underlying physical mechanisms. Leveraging these advantages, we demonstrate accurate engineering of effective potentials in Paul traps governed by the nonlinear Mathieu equations, even in quasi-periodic regimes beyond the limits of weak and fast drive conditions. This approach is feasible with existing setups and holds promise for advancing various applications \cite{kielpinski2002, Leibfried2003, HAFFNER2008, Li2012, Qian2022, xu2023, Liu2023, Andris2025}.

\textit{Method I: Nonlinear Extended Floquet Theory}---We start by introducing the MAFT-HB method. Consider a particle with mass $m$ at position $u(t) \in \mathbb{R}^d$, subjected to a nonlinear potential $U_F(u,t)$, which consists of static and periodically oscillating components. The equation of motion is then given by: 
\begin{equation}
    \ddot{u}(t) =  F(u,t),
    \label{ODF}
\end{equation}
where the dot denotes the time derivative $d/dt$, and $F(u,t) = - \nabla U_{F}(u,t)/m$ is a nonlinear function that oscillates with a period $T = 2 \pi / \Omega$, satisfying $F(u,t) = F(u,t + T)$. The goal is to engineer the specified time-independent effective potential $U_\mathrm{eff}$ by carefully selecting the parameters of $F(u,t)$. 

Due to the periodically oscillating nonlinear terms, the analytical resolution of Eq.~(\ref{ODF}) is generally difficult. The Alternating Frequency/Time Domain Harmonic Balance (AFT-HB) method is a well-established approach for analyzing such nonlinear systems \cite{Cameron1989, Zipu2023, Nicks2024}. This method employs a truncated Fourier series as an ansatz and the Discrete Fourier Transformation (DFT) to convert the nonlinear differential equations into nonlinear algebraic equations. The DFT process eliminates complex algebraic transformations, enabling efficient and accurate analysis even for the non-polynomial equations of motion. 

However, the AFT-HB method cannot be directly employed for Floquet engineering because the aliasing in the DFT process, introduced by the micromotion with the frequency $\Omega/2\pi$ caused by the nonlinear drive and the slower secular motion with $\omega/2\pi$ by $U_\mathrm{eff}$, significantly reduces its accuracy \cite{lindblad2022}. We circumvent this issue with the MAFT-HB method, which employs the Multidimensional Discrete Fourier Transformation (MDFT) that effectively suppresses aliasing and enables efficient, accurate, and general analysis of strongly driven nonlinear systems \cite{Junge2021}.

To analyze the Eq.~(\ref{ODF}) with the MAFT-HB method, we first assume the truncated oscillatory trial function that incorporates both the frequency of the micromotion $\Omega/2\pi$ and the secular motion $\omega/2\pi$ as follows:
\begin{equation}
    u(\xi,\zeta) = \sum_{k = 1}^K \sum_{m = -M}^M \tilde{A}_{mk} e^{-i \left(k \omega \xi + m \Omega \zeta \right)} + \mathrm{c.c}.
    \label{testfunction}
\end{equation}
Here, $\tilde{A}_{mk} = \frac{1}{2}A_{mk} e^{i\theta}$ with $A_{mk} \in \mathbb{R}^d$ where $A_{01}$ represents the amplitude of the secular motion, and $\theta$ is a constant determined by the initial conditions. This trial function represents the multidimensional Fourier expansion with the pseudo-time parameters $\xi$ and $\zeta$ to avoid aliases \cite{Junge2021}. These parameters are set as $\xi=\zeta$ after estimating $A_{mk}$ and $\omega/2\pi$ to calculate the real-time trajectory. Note that the trial function of Eq~(\ref{testfunction}) is different from the conventional truncated Floquet solution \cite{Oka2019}:
\begin{equation}
    u(\xi, \zeta) = e^{-i \omega \xi} \sum_{m = -M}^M \tilde{A}_{m} e^{-i m \Omega \zeta} + \mathrm{c.c},
    \label{floquet}
\end{equation}
where, $\tilde{A}_{m} = \frac{1}{2}A_{m} e^{i \theta}$ with $A_{m} \in \mathbb{R}^d$. The distinction lies in the assumption that in Eq.~(\ref{testfunction}), the secular motion exhibits higher harmonics with frequencies $k\omega/2\pi \ \forall k \in \mathbb{N}^+$ due to the nonlinearity, which is not accounted for in Eq.~(\ref{floquet}) and the Floquet theory. 

As we will see later, Eq.~(\ref{testfunction}) provides a better approximate solution to Eq.~(\ref{ODF}), making it a natural extension of the Floquet theory to the nonlinear differential equations. For clarity, we refer to Eq.~(\ref{floquet}) as the Ordinary Floquet Solution (OFS) and Eq.~(\ref{testfunction}) as the Nonlinear Extended Floquet Solution (NEFS) in this paper.

Next, we discretize $u(\xi,\zeta)$ by sampling it at $\xi = \xi_s = 2 \pi s/\omega M_{\xi} \ \forall s= 1, \ldots, M_{\xi}$ and $\zeta = \zeta_p = 2 \pi p/\Omega M_{\zeta} \ \forall p=1, \ldots, M_{\zeta}$. The functions are then expressed as MDFT vectors:
\begin{align}
    \tilde{u} &= \left[A_{-M1} \ A_{-M2} \ \cdots \ A_{MK} \right]^\top ,\\
    \hat{u} &= \left[u(\xi_1, \zeta_1) \ u(\xi_1, \zeta_2) \ \cdots \ u(\xi_{M_\xi}, \zeta_{M_\zeta}) \right]^\top \notag \\ 
    &= (S \otimes I_d) \tilde{u}, \\
    \hat{F}(\hat{u}) &= \left[F(\hat{u}_1, \zeta_1) \ F(\hat{u}_2, \zeta_2) \  \cdots \ F(\hat{u}_{M_\xi M_\zeta}, \zeta_{M_\zeta}) \right]^\top \notag \\ 
    &= (S \otimes I_d) \tilde{F}(\tilde{u}).
\end{align}
Here, $\top$ denotes transpose, $\tilde{u} = \mathrm{vec}(A)$ with $[A]_{mk} = A_{mk}$ is the vectorized expansion coefficient matrix, $\hat{u}_i \ \forall i=1,\ldots , M_\xi M_\zeta$ is the $i$-th component of $\hat{u}$, $\tilde{F}(\tilde{u})$ is the nonlinear function of $\tilde{u}$, $\otimes$ is the Kronecker product, and $I_d$ is the $d\cross d$ identity matrix. The matrix $S$ is the MDFT operator of dimensions $M_\xi  M_\zeta\cross(2M+1)K$, defined as:
\begin{gather}
    S = 
    \begin{pmatrix}
        \hat{\gamma}_{-M1} & \hat{\gamma}_{-M2} &  \cdots & \hat{\gamma}_{MK} \\
    \end{pmatrix},\\
    \hat{\gamma}_{mk} =  \left[\gamma_{mk}(\xi_1, \zeta_1) \ \gamma_{mk}(\xi_1, \zeta_2) \ \cdots \ \gamma_{mk}(\xi_{M_\xi}, \zeta_{M_\zeta}) \right]^\top,
\end{gather}
and $\gamma_{mk}(\xi_s,\zeta_p) = \frac{1}{2} e^{-i \left(k\omega \xi_s + m \Omega \zeta_p + \theta \right)} + \mathrm{c.c}$. The completeness relation $S S^{-1} = I_{(2M+1)K}$ and the orthogonality of harmonic bases are approximately satisfied when sampling points $M_\xi$ and $M_\zeta$ are sufficiently large.

Finally, we substitute the sampled functions into Eq.~(\ref{ODF}), reducing it to a system of $d(2M+1)K$ nonlinear algebraic equations expressed as:
\begin{align}
    W(\vb*{a};\vb*{b}) &\equiv (S \nabla^2 S^{-1} \otimes I_d) \hat{u} - \hat{F}(\hat{u})  \notag \\
    &= O,
    \label{NSE}
\end{align}
where $[\nabla^2]_{mk} = - \delta_{mk} (k\omega + m\Omega )^2 $. The vector $\vb*{a} = [A_{01},\theta]^\top$ encapsulates parameters determined by initial conditions,  while $\vb*{b} = [A_{mk}, \omega]^\top \ \forall k \neq 1$ represents the parameters estimated by solving  Eq.~(\ref{NSE}). We utilized the $\mathsf{Findroot}$ routine in Mathematica to solve Eq.~(\ref{NSE}), but any appropriate nonlinear solver can be applied. While we focused on the two-fundamental frequency system $(\omega, \Omega)$ for simplicity, this discussion can be extended to systems with more fundamental frequencies, as shown in previous research \cite{Junge2021}.

\textit{Method II: Resolution of Inverse Problems}---We formulate inverse problems by characterizing the effective potential using amplitude-dependent frequency shifts, expressed as follows:
\begin{equation}
    \omega (A,\vb*{\varepsilon}) = \omega_0 \left[1+\sum_{k=2}^{\infty} \varepsilon_k  A^k \right],
    \label{poincare}
\end{equation}
 where $A$ represents the amplitude and $\omega/2\pi$ is the frequency of the secular motion caused by the effective potential $U_\mathrm{eff} = \frac{1}{2} m \omega_0^2 (u^2 + \sum_{k=3}^{\infty} C_k u^k)$. Here, expansion coefficients $\vb*{\varepsilon} = [\varepsilon_2 \ \varepsilon_3 \ \cdots]^\top$ are directly related to the anharmonicity coefficients $C_k$ as shown in \cite{Doroudi2009, Goldman2010}. 

Assuming $F(u,t)$ is dependent on control parameters $\alpha_i \ \forall i = 1,\ldots, N_c$, the inverse problem correspond to optimizing $[\vb*{\alpha}]_i = \alpha_i $ to achieve the target $U_\mathrm{eff}$ with specified anharmonicities $C_k$ under the constraint of Eq.~(\ref{NSE}) and Eq.~(\ref{poincare}). We address this constrained optimization problem by incorporating the constraints and additional free parameters into the equation, emulating the Lagrange multiplier method \cite{Rockafellar1993}. The problem is then reformulated as a set of simultaneous nonlinear algebraic equations:
\begin{align}
    T(\vb*{a}_0, \vb*{a}_1,  \cdots , \vb*{a}_{N_c}; \vb*{b}_0,\vb*{b}_1,  \cdots , \vb*{b}_{N_c}) &\equiv 
    \begin{pmatrix}
        W(\vb*{a}_0;\vb*{b}_0) \\
        W(\vb*{a}_1;\vb*{b}_1) \\
        \vdots \\
        W(\vb*{a}_{N_c};\vb*{b}_{N_c})
    \end{pmatrix} \notag \\
    &= O,
    \label{tensor}
\end{align}
where $W(\vb*{a}_j;\vb*{b}_j)$ is expressed using trial functions:
\begin{align}
u_j(\xi, \zeta) =  \sum_{k=1}^{K}\sum_{m=-M}^{M}\tilde{A}_{mk}^{(j)}e^{- i\left(k \omega \xi + m \Omega \zeta \right)} + \mathrm{c.c}.
\label{trial2}
\end{align}
Here, $\tilde{A}_{mk}^{(j)} = \frac{1}{2}A_{mk}^{(j)} e^{i \theta}$ with $A_{mk}^{(j)} \in \mathbb{R}^d $ for $j = 0,1,\ldots N_c$. The input parameters are defined as $\vb*{a}_j = [A_{01}^{(j)}, \theta]^\top$, and the output parameters as $\vb*{b}_j = [A_{mk}^{(j)}, \omega(A_{01}^{(j)}, \vb*{\varepsilon}), \vb*{\alpha}]^\top \ \forall k \neq 1$. Eq.~(\ref{tensor}) establish the nonlinear relationships between $A_{mk}^{(j)}$s and $\vb*{\alpha}$, imposing the specified amplitude-frequency relations $\omega(A_{01}^{(j)},\vb*{\varepsilon})$ for $A_{01}^{(j)} \ \forall j = 0,1,\ldots, N_c$. Therefore, the inverse problem is reduced to solving Eq.~(\ref{tensor}) to determine the optimal control parameters $\vb*{\alpha}$ that realize the amplitude-frequency relations $\omega(A_{01}^{(j)},\vb*{\varepsilon})$ specified by the target effective potential $U_\mathrm{eff}$ through Eq.~(\ref{poincare}).

It is worth comparing the number of variables to the number of constraints in Eq.~(\ref{tensor}). Given the initial conditions $A_{01}^{(j)}$ and $\theta$, the remaining variables are $\vb*{b}_0, \ldots \vb*{b}_{N_{c}}$ comprising $d(N_c+1) (2M+1) K$ free parameters. In contrast, the constraints $W(\vb*{a}_j ; \vb*{b}_j) = O \ \forall j=0,\ldots,N_c$ contain $d(2M+1)K$ equations each. Therefore, the total number of constraints is also $d(N_c+1) (2M+1)K$. Thus, the Eq.~(\ref{tensor}) is well-determined and can be solved using root-finding algorithms \cite{Tarantola2005}.

\textit{Example: Paul Traps}--- Suppressing anharmonicity offers significant advantages in Paul traps across various applications. Anharmonic trap potentials decrease mass spectrometry resolution \cite{Liu2023}, reduce the precision of laser-free single-spin measurement \cite{Peng2017, Qian2022}, limit the sensitivity of ion interferometers \cite{campbell2017, Shinjo2021}, lowers sympathetic cooling efficiency \cite{Guggemos2015, Francisco2017, Bohman2021}, and degrades the fidelity of two-qubit gates \cite{Home2011, Sutherland2022}. Conversely, anharmonicities can also be valuable resources for generating nonclassical states of ion motion, such as the Gottesman-Kitaev-Preskill state, essential building blocks of quantum computations \cite{Gottesman2001, Flühmann2019, fluhmann2019, Rojkov2024, Leon2024}. Therefore, anharmonicity modulation is of great importance in Paul traps.

Based on the present method, we propose an approach for engineering the effective potential of Paul traps using static electric potentials. This has been challenging due to the complexity of the nonlinear Mathieu equations, restricting implementations to the small $q$ parameters regime $(q \ll 0.1)$, where perturbation theory applies \cite{Home2011}. In contrast, our semi-analytical engineering method enables precise tuning of anharmonicities even in the large $q$ parameter regime $(q > 0.5)$. This allows for accurate spectroscopic measurements across a broader range of parameters and facilitates quantum operations with faster execution, reduced motional heating, and simpler cooling requirements \cite{xu2023}. Experimentally, this technique requires only tailored static electric fields, which have already been implemented in various setups \cite{Schmied2009, Kim2010, romaszko2020, Palani2023, Seidling2024, Lysne2024}.

We consider a one-dimensional Paul trap $(d = 1)$ using a periodically oscillating nonlinear potential $U_{\mathrm{ac}}(u,t) = V_\mathrm{ac} \cos{(\Omega t)} \sum_{k=2}^{\infty} C_{k}^{\mathrm{ac}} u^{k}$ and the nonlinear static potential $U_{\mathrm{dc}}(u) = \sum_{i=1}^{N_e} \sum_{k=2}^{\infty} V_\mathrm{dc}^{(i)} C_{k i}^{\mathrm{dc}} u^{k}$ generated by $N_e$ compensation electrodes. The net potential is then given by $U_{F}(u,t) = U_{\mathrm{ac}}(u,t) + U_{\mathrm{dc}}(u)$, resulting in the equation of motion for a charged particle with mass $M$ and charge $Q$, known as the nonlinear Mathieu equation:
\begin{align}
    \frac{d^2 u}{{d\xi}^2} - 2 q &\cos{2 \xi} \left(u + \frac{1}{2}\sum_{k=3}^{\infty} k \alpha_{k}^{\mathrm{ac}} u^{k-1} \right) \notag \\
    &+ a \left( u + \frac{1}{2} \sum_{k=3}^{\infty} k \alpha_{k}^{\mathrm{dc}} u^{k-1} \right) = 0.
    \label{duffing}
\end{align}
Here, we use the normalized time $\xi = \Omega t /2$ and define the associated parameters as $q = - 4QV_\mathrm{ac}C_{2}^{\mathrm{ac}}/M \Omega^2$ the Mathieu $q$ parameter, $\alpha_{k}^{\mathrm{ac}} = C_{k}^{\mathrm{ac}}/C_{2}^{\mathrm{ac}}$ the normalized $k$-th nonlinearity coefficient of the AC potential, $a = -2q \sum_{i=1}^{N_e} C_{2i}^{\mathrm{dc}} V_\mathrm{dc}^{(i)}/C_{2}^{\mathrm{ac}} V_\mathrm{ac}$ the Mathieu $a$ parameter, and $\alpha_{k}^{\mathrm{dc}} = \sum_{i=1}^{N_e} C_{ki}^{\mathrm{dc}} V_\mathrm{dc}^{(i)} / \sum_{i=1}^{N_e} C_{2i}^{\mathrm{dc}} V_\mathrm{dc}^{(i)}$ the normalized $k$-th nonlinearity coefficient of the DC potential. In this paper, we assume parity symmetry at $u = 0$, so that odd nonlinear terms vanish $(\alpha_{2k-1}^{\mathrm{ac}} = \alpha_{2k-1}^{\mathrm{dc}} = 0 \ \forall k \in \mathbb{N})$, and introduce $\tilde{\alpha}_k^{\mathrm{dc}} = a \alpha_k^{\mathrm{dc}} = -2q \sum_{i=1}^{N_e} C_{ki}^{\mathrm{dc}} V_\mathrm{dc}^{(i)} / C_{2}^{\mathrm{ac}} V_\mathrm{ac}$ for simplicity.

We first solved Eq.~(\ref{duffing}) using MAFT-HB method, assuming the parameters $q = 0.7$, $\alpha_4^{\mathrm{ac}} = -0.2$, $\alpha_6^{\mathrm{ac}} = -0.4$, $\alpha_8^{\mathrm{ac}} = 0.01$, $\alpha_{k}^{\mathrm{ac}} = 0 \ \forall k \ge 9$, and no DC potentials $ (a=\tilde{\alpha}_k^{\mathrm{dc}} = 0 \ \forall k \ge 3)$, with initial conditions $A_{01} = 0.2$ and $\theta = 0$. As trial functions, we chose NEFS [Eq.~(\ref{testfunction})] truncated at $m = \pm 7$ and $k = 8$ and OFS [Eq.~(\ref{floquet})] at $m = \pm 7$ with the sampling points of $M_{\xi} = M_{\zeta} = 15$. Figure~\ref{Fig1} compared the results of NEFS (blue), OFS (orange), and the numerical integration of Eq.~(\ref{duffing}) using the $8$-th order Runge-Kutta method (dashed black). The normalized difference $|\Delta u (\xi)/u(0)|$, which quantifies deviations in $u(\xi)$ between numerical integration and the MAFT-HB methods, demonstrated that NEFS accurately reproduces the numerical results while OFS shows a large discrepancy. As expected, the Floquet theory-based OFS fails to describe the trajectory of Eq.~(\ref{duffing}), confirming that the NEFS is a natural extension of Floquet theory to nonlinear differential equations.

\begin{figure}[t]
\includegraphics[width=8.4cm]{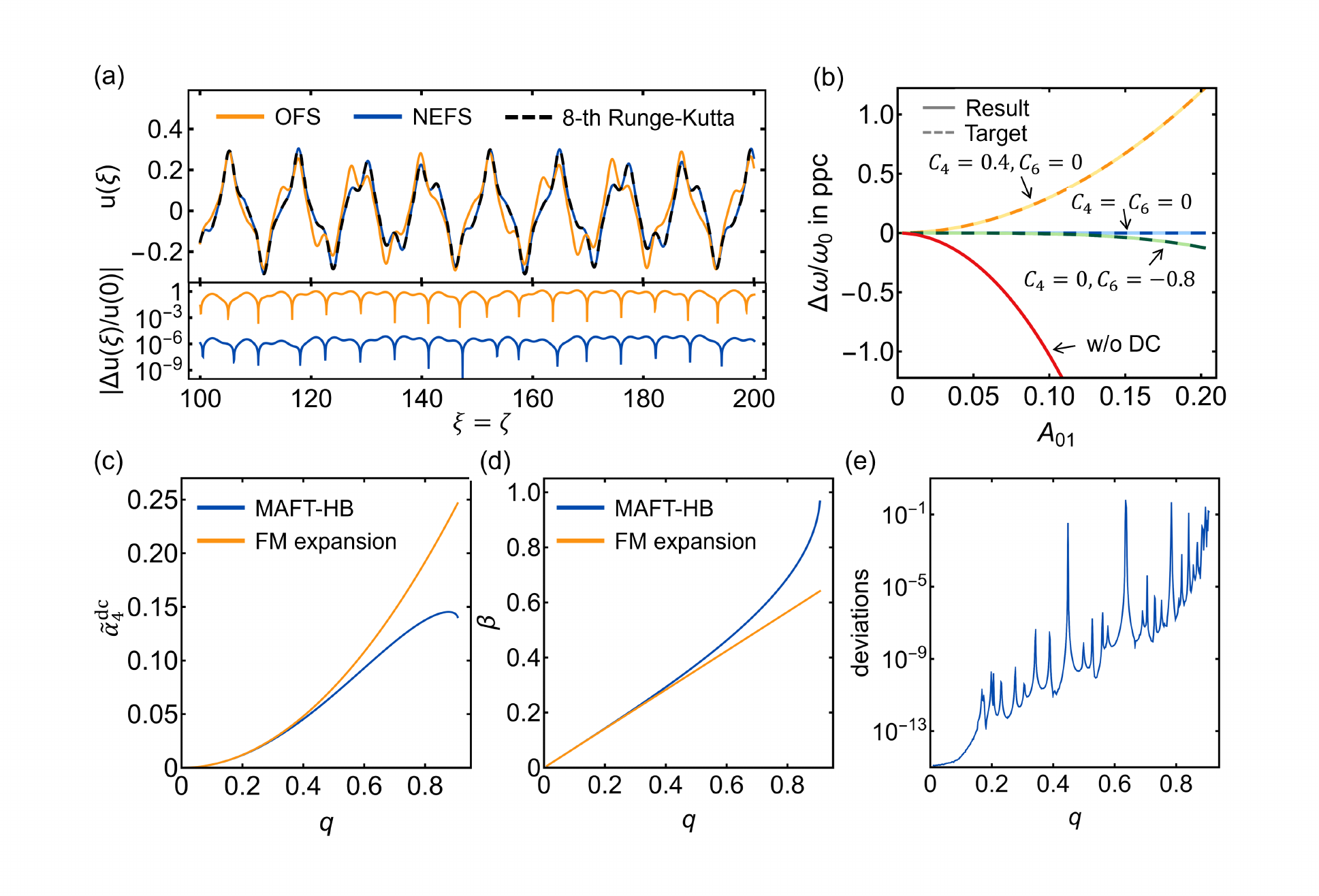}
\caption{\label{Fig1} Demonstration of semi-analytical engineering in Paul traps. (a) The trajectory of a trapped particle using the $8$-th Runge-Kutta method (dashed black) compared with results of the MAFT-HB method using OFS (orange) and NEFS (blue), along with their deviations (lower panel). (b) Relative amplitude-dependent frequency shifts of the target effective potentials (dashed lines), the engineered potentials (solid lines), and the shifts without DC compensation (red). A ppc is $10^{-2}$. (c-e) Optimized DC anharmonicity $\tilde{\alpha}_4^{\mathrm{dc}}$ to achieve $C_4 = 0.4$, corresponding normalized secular frequency $\beta = 2\omega/\Omega$, and the maximum deviations from the numerical integration result as functions of the Mathieu $q$ parameters.}
\end{figure}

We next addressed the inverse problem for engineering the effective potential of Paul traps. Here, nonlinear DC potentials were superposed onto the AC potential assumed in Fig.~\ref{Fig1}(a), using four compensation electrodes $(N_e = 4)$. This enabled us to select the control parameters $\left[\tilde{\alpha}_4^{\mathrm{dc}}, \tilde{\alpha}_6^{\mathrm{dc}}, \tilde{\alpha}_8^{\mathrm{dc}} \right]^\top = \vb*{\alpha}$ on the condition $a = 0$. Control parameters were optimized by solving $T(\vb*{a}_0, \vb*{a}_1, \vb*{a}_2, \vb*{a}_{3}; \vb*{b}_0, \vb*{b}_1,\vb*{b}_2, \vb*{b}_{3}) = O$ with $A_{01}^{(0)} = 10^{-5}, A_{01}^{(1)} = 10^{-4}, A_{01}^{(2)} = 10^{-3}, A_{01}^{(3)} = 10^{-2}$ to engineer anharmonicities $C_4 = 0, C_6 = -0.8, C_8 = 0$ (green), $C_4 = C_6 = C_8 = 0$ (blue), and $C_4 = 0.4, C_6 = C_8 = 0$ (orange) of the target effective potentials. Their corresponding relative amplitude-dependent frequency shifts $\Delta \omega / \omega_0 = [\omega(A_{01}, \vb*{\varepsilon}) - \omega_0]/\omega_0$ are shown as dashed lines in Fig.~\ref{Fig1}(b), along with the original shifts without DC potentials (red). The resulting optimized frequency shifts (solid lines) closely matched the targets, with errors dominated by $|\Delta u (\xi)/u(0)|$, confirming the successful resolution of the inverse problem.

Finally, we compared our method to the perturbative approach used in standard Floquet engineering \cite{higashikawa2018}. In this method, the equation of motion is expressed as $\dot{\vb*{\phi}} = \vb*{F}(\vb*{\phi},t)$, where $\vb*{\phi} = [u,\dot{u}]^\top$ and $\vb*{F}(\vb*{\phi},t) = [\dot{u}, F(u,t)]^\top$, and the driving force $\vb*{F}(\vb*{\phi},t)$ is expanded as $\vb*{F}(\vb*{\phi},t) = \sum_m \vb*{F}_m (\vb*{\phi}) e^{-im\Omega t}$. Applying Floquet theory and the Floquet-Magnus (FM) expansion to the corresponding Liouville's equation yields the effective equation of motion \cite{higashikawa2018}:
\begin{align}
    \dot{\vb*{\phi}} &= \vb*{F}_\mathrm{eff}(\vb*{\phi}) \notag \\
     &= \vb*{F}_0(\vb*{\phi}) + \sum_{m\neq 0} \frac{ [\vb*{F}_{-m},\vb*{F}_m]}{2m\Omega} \notag \\
    &+ \sum_{m \neq 0} \left(\frac{[\vb*{F}_{-m}, [\vb*{F}_0, \vb*{F}_m]]}{2 (m\Omega)^2} + \sum_{n\neq0,m} \frac{[\vb*{F}_{-n}, [\vb*{F}_{n-m},\vb*{F}_m]]}{3nm  \Omega^2} \right) \notag \\
    &+  \mathcal{O}(\Omega^{-3}).
    \label{Magnus}
\end{align}
Here, $[\cdot,\cdot ]$ is the Poisson bracket, and $\vb*{F}_\mathrm{eff}(\vb*{\phi}) = [\dot{u}, F_\mathrm{eff}(u)]^\top$ is the time-independent effective force. The effective potential is determined by $F_\mathrm{eff}(u) = - \nabla U_\mathrm{eff}(u)$. 

We solved $T(\vb*{a}_0, \vb*{a}_1; \vb*{b}_0, \vb*{b}_1) = O$ with the control parameter $\tilde{\alpha}_4^{\mathrm{dc}}$ to achieve $C_4 = 0.4$, and compared the optimized results, including their normalized secular frequencies $\beta = 2 \omega/\Omega$, with those predicted by the second-order truncation of Eq.~(\ref{Magnus}) in Fig.~\ref{Fig1}(c,d). Significant discrepancy arose between the two approaches at the higher $q$, where the weak and fast drive assumptions for the FM expansion break down. To verify the accuracy of our approach, we plotted the maximum deviations of $|\Delta u (\xi)/u(0)|$ over $0<\xi=\zeta<200$ in Fig.~\ref{Fig1}(e). Although nonlinear parametric excitations introduce higher-order harmonics not accounted for in the used NEFS and increase the deviations at certain $q$, their overall values are small even in the high $q$ regime where quasi-periodic oscillations dominate. This demonstrates the robustness and accuracy of our method across a wide range of $q$, overcoming the limitations of the perturbative approach.

\textit{Conclusion and Outlook}---We have proposed a nonperturbative, semi-analytical framework to solve inverse problems in strongly driven nonlinear systems. Building on the MAFT-HB method, we extended the Floquet theory to nonlinear differential equations and established a novel constrained optimization technique inspired by the Lagrange multiplier method. This method enables precise engineering of effective potentials, demonstrated in Paul traps, providing significant accuracy improvements over the standard Floquet engineering framework in regimes where weak and fast drive conditions break down. Our framework is applicable to various periodically driven nonlinear systems, including driven optical lattices \cite{Yin2022}, micromagnets \cite{Gilbert2004}, Josephson junctions \cite{Pietik2019, Frattini2018}, and quantum dots \cite{bahar2022}. For instance, we applied the present method to driven optical lattices in the supplemental material, suggesting its potential to enhance resolved-sideband cooling and diversify bosonic quantum simulations.

The present method can naturally extend to the many-body and quasi-periodically driven systems by adding more frequency components with which MDFT is compatible \cite{Junge2021}. It holds potential for solving other complex optimization challenges, such as heating-suppressed shuttling at Paul trap X-junctions \cite{blakestad2009, sterk2022, delaney2024}. Future advancements could include introducing state-of-the-art Harmonic Balance methods for accurately analyzing and optimizing systems under noise or stochastic drives \cite{Didier2012, Didier2013, Miguel2022, Chao2023}. Extending the trial function to general orthogonal polynomial function bases beyond harmonics would enable applications of the present method to more complex systems, such as optical Kerr cavities, waveguides, and Bose-Einstein condensates governed by nonlinear Schr\"{o}dinger equations \cite{Goldman2023, gesla2024}. Furthermore, incorporating noncommutative finite transformation (NcFT) could pave the way for non-perturbative, semi-analytical engineering of more general quantum systems \cite{Guo2024, Xu2025}. 

\textit{Acknowledgements}---The authors thank Ippei Nakamura for fruitful discussions. This research was supported by JST SPRING (Grant No.JPMJSP2108) and JST PRESTO (Grant No.JPMJPR2258).

\bibliography{apssamp}

\begin{thebibliography}{83}%
\makeatletter
\providecommand \@ifxundefined [1]{%
 \@ifx{#1\undefined}
}%
\providecommand \@ifnum [1]{%
 \ifnum #1\expandafter \@firstoftwo
 \else \expandafter \@secondoftwo
 \fi
}%
\providecommand \@ifx [1]{%
 \ifx #1\expandafter \@firstoftwo
 \else \expandafter \@secondoftwo
 \fi
}%
\providecommand \natexlab [1]{#1}%
\providecommand \enquote  [1]{``#1''}%
\providecommand \bibnamefont  [1]{#1}%
\providecommand \bibfnamefont [1]{#1}%
\providecommand \citenamefont [1]{#1}%
\providecommand \href@noop [0]{\@secondoftwo}%
\providecommand \href [0]{\begingroup \@sanitize@url \@href}%
\providecommand \@href[1]{\@@startlink{#1}\@@href}%
\providecommand \@@href[1]{\endgroup#1\@@endlink}%
\providecommand \@sanitize@url [0]{\catcode `\\12\catcode `\$12\catcode `\&12\catcode `\#12\catcode `\^12\catcode `\_12\catcode `\%12\relax}%
\providecommand \@@startlink[1]{}%
\providecommand \@@endlink[0]{}%
\providecommand \url  [0]{\begingroup\@sanitize@url \@url }%
\providecommand \@url [1]{\endgroup\@href {#1}{\urlprefix }}%
\providecommand \urlprefix  [0]{URL }%
\providecommand \Eprint [0]{\href }%
\providecommand \doibase [0]{https://doi.org/}%
\providecommand \selectlanguage [0]{\@gobble}%
\providecommand \bibinfo  [0]{\@secondoftwo}%
\providecommand \bibfield  [0]{\@secondoftwo}%
\providecommand \translation [1]{[#1]}%
\providecommand \BibitemOpen [0]{}%
\providecommand \bibitemStop [0]{}%
\providecommand \bibitemNoStop [0]{.\EOS\space}%
\providecommand \EOS [0]{\spacefactor3000\relax}%
\providecommand \BibitemShut  [1]{\csname bibitem#1\endcsname}%
\let\auto@bib@innerbib\@empty
\bibitem [{\citenamefont {Winterfeldt}\ \emph {et~al.}(2008)\citenamefont {Winterfeldt}, \citenamefont {Spielmann},\ and\ \citenamefont {Gerber}}]{Winterfeldt2008}%
  \BibitemOpen
  \bibfield  {author} {\bibinfo {author} {\bibfnamefont {C.}~\bibnamefont {Winterfeldt}}, \bibinfo {author} {\bibfnamefont {C.}~\bibnamefont {Spielmann}},\ and\ \bibinfo {author} {\bibfnamefont {G.}~\bibnamefont {Gerber}},\ }\bibfield  {title} {\bibinfo {title} {Colloquium: Optimal control of high-harmonic generation},\ }\href@noop {} {\bibfield  {journal} {\bibinfo  {journal} {Rev. Mod. Phys.}\ }\textbf {\bibinfo {volume} {80}},\ \bibinfo {pages} {117} (\bibinfo {year} {2008})}\BibitemShut {NoStop}%
\bibitem [{\citenamefont {Zaletel}\ \emph {et~al.}(2023)\citenamefont {Zaletel}, \citenamefont {Lukin}, \citenamefont {Monroe}, \citenamefont {Nayak}, \citenamefont {Wilczek},\ and\ \citenamefont {Yao}}]{Zaletel2023}%
  \BibitemOpen
  \bibfield  {author} {\bibinfo {author} {\bibfnamefont {M.~P.}\ \bibnamefont {Zaletel}}, \bibinfo {author} {\bibfnamefont {M.}~\bibnamefont {Lukin}}, \bibinfo {author} {\bibfnamefont {C.}~\bibnamefont {Monroe}}, \bibinfo {author} {\bibfnamefont {C.}~\bibnamefont {Nayak}}, \bibinfo {author} {\bibfnamefont {F.}~\bibnamefont {Wilczek}},\ and\ \bibinfo {author} {\bibfnamefont {N.~Y.}\ \bibnamefont {Yao}},\ }\bibfield  {title} {\bibinfo {title} {Colloquium: Quantum and classical discrete time crystals},\ }\href@noop {} {\bibfield  {journal} {\bibinfo  {journal} {Rev. Mod. Phys.}\ }\textbf {\bibinfo {volume} {95}},\ \bibinfo {pages} {031001} (\bibinfo {year} {2023})}\BibitemShut {NoStop}%
\bibitem [{\citenamefont {Eckardt}(2017)}]{Eckardt2017}%
  \BibitemOpen
  \bibfield  {author} {\bibinfo {author} {\bibfnamefont {A.}~\bibnamefont {Eckardt}},\ }\bibfield  {title} {\bibinfo {title} {Colloquium: Atomic quantum gases in periodically driven optical lattices},\ }\href@noop {} {\bibfield  {journal} {\bibinfo  {journal} {Rev. Mod. Phys.}\ }\textbf {\bibinfo {volume} {89}},\ \bibinfo {pages} {011004} (\bibinfo {year} {2017})}\BibitemShut {NoStop}%
\bibitem [{\citenamefont {Hensinger}\ \emph {et~al.}(2001)\citenamefont {Hensinger}, \citenamefont {H{\"a}ffner}, \citenamefont {Browaeys}, \citenamefont {Heckenberg}, \citenamefont {Helmerson}, \citenamefont {McKenzie}, \citenamefont {Milburn}, \citenamefont {Phillips}, \citenamefont {Rolston}, \citenamefont {Rubinsztein-Dunlop},\ and\ \citenamefont {Upcroft}}]{Hensinger2001}%
  \BibitemOpen
  \bibfield  {author} {\bibinfo {author} {\bibfnamefont {W.~K.}\ \bibnamefont {Hensinger}}, \bibinfo {author} {\bibfnamefont {H.}~\bibnamefont {H{\"a}ffner}}, \bibinfo {author} {\bibfnamefont {A.}~\bibnamefont {Browaeys}}, \bibinfo {author} {\bibfnamefont {N.~R.}\ \bibnamefont {Heckenberg}}, \bibinfo {author} {\bibfnamefont {K.}~\bibnamefont {Helmerson}}, \bibinfo {author} {\bibfnamefont {C.}~\bibnamefont {McKenzie}}, \bibinfo {author} {\bibfnamefont {G.~J.}\ \bibnamefont {Milburn}}, \bibinfo {author} {\bibfnamefont {W.~D.}\ \bibnamefont {Phillips}}, \bibinfo {author} {\bibfnamefont {S.~L.}\ \bibnamefont {Rolston}}, \bibinfo {author} {\bibfnamefont {H.}~\bibnamefont {Rubinsztein-Dunlop}},\ and\ \bibinfo {author} {\bibfnamefont {B.}~\bibnamefont {Upcroft}},\ }\bibfield  {title} {\bibinfo {title} {Dynamical tunnelling of ultracold atoms},\ }\href@noop {} {\bibfield  {journal} {\bibinfo  {journal} {Nature}\ }\textbf {\bibinfo {volume} {412}},\ \bibinfo {pages} {52} (\bibinfo {year} {2001})}\BibitemShut {NoStop}%
\bibitem [{\citenamefont {Choi}\ \emph {et~al.}(2020)\citenamefont {Choi}, \citenamefont {Zhou}, \citenamefont {Knowles}, \citenamefont {Landig}, \citenamefont {Choi},\ and\ \citenamefont {Lukin}}]{Choi2020}%
  \BibitemOpen
  \bibfield  {author} {\bibinfo {author} {\bibfnamefont {J.}~\bibnamefont {Choi}}, \bibinfo {author} {\bibfnamefont {H.}~\bibnamefont {Zhou}}, \bibinfo {author} {\bibfnamefont {H.~S.}\ \bibnamefont {Knowles}}, \bibinfo {author} {\bibfnamefont {R.}~\bibnamefont {Landig}}, \bibinfo {author} {\bibfnamefont {S.}~\bibnamefont {Choi}},\ and\ \bibinfo {author} {\bibfnamefont {M.~D.}\ \bibnamefont {Lukin}},\ }\bibfield  {title} {\bibinfo {title} {Robust dynamic hamiltonian engineering of many-body spin systems},\ }\href@noop {} {\bibfield  {journal} {\bibinfo  {journal} {Phys. Rev. X}\ }\textbf {\bibinfo {volume} {10}},\ \bibinfo {pages} {031002} (\bibinfo {year} {2020})}\BibitemShut {NoStop}%
\bibitem [{\citenamefont {Nguyen}\ \emph {et~al.}(2024)\citenamefont {Nguyen}, \citenamefont {Kim}, \citenamefont {Hashim}, \citenamefont {Goss}, \citenamefont {Marinelli}, \citenamefont {Bhandari}, \citenamefont {Das}, \citenamefont {Naik}, \citenamefont {Kreikebaum}, \citenamefont {Jordan} \emph {et~al.}}]{nguyen2024}%
  \BibitemOpen
  \bibfield  {author} {\bibinfo {author} {\bibfnamefont {L.~B.}\ \bibnamefont {Nguyen}}, \bibinfo {author} {\bibfnamefont {Y.}~\bibnamefont {Kim}}, \bibinfo {author} {\bibfnamefont {A.}~\bibnamefont {Hashim}}, \bibinfo {author} {\bibfnamefont {N.}~\bibnamefont {Goss}}, \bibinfo {author} {\bibfnamefont {B.}~\bibnamefont {Marinelli}}, \bibinfo {author} {\bibfnamefont {B.}~\bibnamefont {Bhandari}}, \bibinfo {author} {\bibfnamefont {D.}~\bibnamefont {Das}}, \bibinfo {author} {\bibfnamefont {R.~K.}\ \bibnamefont {Naik}}, \bibinfo {author} {\bibfnamefont {J.~M.}\ \bibnamefont {Kreikebaum}}, \bibinfo {author} {\bibfnamefont {A.~N.}\ \bibnamefont {Jordan}}, \emph {et~al.},\ }\bibfield  {title} {\bibinfo {title} {Programmable heisenberg interactions between floquet qubits},\ }\href@noop {} {\bibfield  {journal} {\bibinfo  {journal} {Nature Physics}\ }\textbf {\bibinfo {volume} {20}},\ \bibinfo {pages} {240} (\bibinfo {year} {2024})}\BibitemShut {NoStop}%
\bibitem [{\citenamefont {Kiefer}\ \emph {et~al.}(2019)\citenamefont {Kiefer}, \citenamefont {Hakelberg}, \citenamefont {Wittemer}, \citenamefont {Berm\'udez}, \citenamefont {Porras}, \citenamefont {Warring},\ and\ \citenamefont {Schaetz}}]{Kiefer2019}%
  \BibitemOpen
  \bibfield  {author} {\bibinfo {author} {\bibfnamefont {P.}~\bibnamefont {Kiefer}}, \bibinfo {author} {\bibfnamefont {F.}~\bibnamefont {Hakelberg}}, \bibinfo {author} {\bibfnamefont {M.}~\bibnamefont {Wittemer}}, \bibinfo {author} {\bibfnamefont {A.}~\bibnamefont {Berm\'udez}}, \bibinfo {author} {\bibfnamefont {D.}~\bibnamefont {Porras}}, \bibinfo {author} {\bibfnamefont {U.}~\bibnamefont {Warring}},\ and\ \bibinfo {author} {\bibfnamefont {T.}~\bibnamefont {Schaetz}},\ }\bibfield  {title} {\bibinfo {title} {Floquet-engineered vibrational dynamics in a two-dimensional array of trapped ions},\ }\href@noop {} {\bibfield  {journal} {\bibinfo  {journal} {Phys. Rev. Lett.}\ }\textbf {\bibinfo {volume} {123}},\ \bibinfo {pages} {213605} (\bibinfo {year} {2019})}\BibitemShut {NoStop}%
\bibitem [{\citenamefont {Yin}\ \emph {et~al.}(2022)\citenamefont {Yin}, \citenamefont {Lu}, \citenamefont {Li}, \citenamefont {Xia}, \citenamefont {Wang}, \citenamefont {Zhang},\ and\ \citenamefont {Chang}}]{Yin2022}%
  \BibitemOpen
  \bibfield  {author} {\bibinfo {author} {\bibfnamefont {M.-J.}\ \bibnamefont {Yin}}, \bibinfo {author} {\bibfnamefont {X.-T.}\ \bibnamefont {Lu}}, \bibinfo {author} {\bibfnamefont {T.}~\bibnamefont {Li}}, \bibinfo {author} {\bibfnamefont {J.-J.}\ \bibnamefont {Xia}}, \bibinfo {author} {\bibfnamefont {T.}~\bibnamefont {Wang}}, \bibinfo {author} {\bibfnamefont {X.-F.}\ \bibnamefont {Zhang}},\ and\ \bibinfo {author} {\bibfnamefont {H.}~\bibnamefont {Chang}},\ }\bibfield  {title} {\bibinfo {title} {Floquet engineering hz-level rabi spectra in shallow optical lattice clock},\ }\href@noop {} {\bibfield  {journal} {\bibinfo  {journal} {Phys. Rev. Lett.}\ }\textbf {\bibinfo {volume} {128}},\ \bibinfo {pages} {073603} (\bibinfo {year} {2022})}\BibitemShut {NoStop}%
\bibitem [{\citenamefont {Guo}\ and\ \citenamefont {Peano}(2024)}]{Guo2024}%
  \BibitemOpen
  \bibfield  {author} {\bibinfo {author} {\bibfnamefont {L.}~\bibnamefont {Guo}}\ and\ \bibinfo {author} {\bibfnamefont {V.}~\bibnamefont {Peano}},\ }\bibfield  {title} {\bibinfo {title} {Engineering arbitrary hamiltonians in phase space},\ }\href@noop {} {\bibfield  {journal} {\bibinfo  {journal} {Phys. Rev. Lett.}\ }\textbf {\bibinfo {volume} {132}},\ \bibinfo {pages} {023602} (\bibinfo {year} {2024})}\BibitemShut {NoStop}%
\bibitem [{\citenamefont {Oka}\ and\ \citenamefont {Kitamura}(2019)}]{Oka2019}%
  \BibitemOpen
  \bibfield  {author} {\bibinfo {author} {\bibfnamefont {T.}~\bibnamefont {Oka}}\ and\ \bibinfo {author} {\bibfnamefont {S.}~\bibnamefont {Kitamura}},\ }\bibfield  {title} {\bibinfo {title} {Floquet engineering of quantum materials},\ }\href@noop {} {\bibfield  {journal} {\bibinfo  {journal} {Annual Review of Condensed Matter Physics}\ }\textbf {\bibinfo {volume} {10}},\ \bibinfo {pages} {387} (\bibinfo {year} {2019})}\BibitemShut {NoStop}%
\bibitem [{\citenamefont {Goldman}\ and\ \citenamefont {Dalibard}(2014)}]{Goldman2014}%
  \BibitemOpen
  \bibfield  {author} {\bibinfo {author} {\bibfnamefont {N.}~\bibnamefont {Goldman}}\ and\ \bibinfo {author} {\bibfnamefont {J.}~\bibnamefont {Dalibard}},\ }\bibfield  {title} {\bibinfo {title} {Periodically driven quantum systems: Effective hamiltonians and engineered gauge fields},\ }\href@noop {} {\bibfield  {journal} {\bibinfo  {journal} {Phys. Rev. X}\ }\textbf {\bibinfo {volume} {4}},\ \bibinfo {pages} {031027} (\bibinfo {year} {2014})}\BibitemShut {NoStop}%
\bibitem [{\citenamefont {Kovacic}\ \emph {et~al.}(2018)\citenamefont {Kovacic}, \citenamefont {Rand},\ and\ \citenamefont {Mohamed~Sah}}]{kovacic2018}%
  \BibitemOpen
  \bibfield  {author} {\bibinfo {author} {\bibfnamefont {I.}~\bibnamefont {Kovacic}}, \bibinfo {author} {\bibfnamefont {R.}~\bibnamefont {Rand}},\ and\ \bibinfo {author} {\bibfnamefont {S.}~\bibnamefont {Mohamed~Sah}},\ }\bibfield  {title} {\bibinfo {title} {Mathieu's equation and its generalizations: Overview of stability charts and their features},\ }\href@noop {} {\bibfield  {journal} {\bibinfo  {journal} {Applied Mechanics Reviews}\ }\textbf {\bibinfo {volume} {70}},\ \bibinfo {pages} {020802} (\bibinfo {year} {2018})}\BibitemShut {NoStop}%
\bibitem [{\citenamefont {Stamper-Kurn}\ and\ \citenamefont {Ueda}(2013)}]{Stamper2013}%
  \BibitemOpen
  \bibfield  {author} {\bibinfo {author} {\bibfnamefont {D.~M.}\ \bibnamefont {Stamper-Kurn}}\ and\ \bibinfo {author} {\bibfnamefont {M.}~\bibnamefont {Ueda}},\ }\bibfield  {title} {\bibinfo {title} {Spinor bose gases: Symmetries, magnetism, and quantum dynamics},\ }\href {https://doi.org/10.1103/RevModPhys.85.1191} {\bibfield  {journal} {\bibinfo  {journal} {Rev. Mod. Phys.}\ }\textbf {\bibinfo {volume} {85}},\ \bibinfo {pages} {1191} (\bibinfo {year} {2013})}\BibitemShut {NoStop}%
\bibitem [{\citenamefont {Gilbert}(2004)}]{Gilbert2004}%
  \BibitemOpen
  \bibfield  {author} {\bibinfo {author} {\bibfnamefont {T.}~\bibnamefont {Gilbert}},\ }\bibfield  {title} {\bibinfo {title} {A phenomenological theory of damping in ferromagnetic materials},\ }\href@noop {} {\bibfield  {journal} {\bibinfo  {journal} {IEEE Transactions on Magnetics}\ }\textbf {\bibinfo {volume} {40}},\ \bibinfo {pages} {3443} (\bibinfo {year} {2004})}\BibitemShut {NoStop}%
\bibitem [{\citenamefont {Marin~Bukov}\ and\ \citenamefont {Polkovnikov}(2015)}]{Martin2015}%
  \BibitemOpen
  \bibfield  {author} {\bibinfo {author} {\bibfnamefont {L.~D.}\ \bibnamefont {Marin~Bukov}}\ and\ \bibinfo {author} {\bibfnamefont {A.}~\bibnamefont {Polkovnikov}},\ }\bibfield  {title} {\bibinfo {title} {Universal high-frequency behavior of periodically driven systems: from dynamical stabilization to floquet engineering},\ }\href@noop {} {\bibfield  {journal} {\bibinfo  {journal} {Advances in Physics}\ }\textbf {\bibinfo {volume} {64}},\ \bibinfo {pages} {139} (\bibinfo {year} {2015})}\BibitemShut {NoStop}%
\bibitem [{\citenamefont {Higashikawa}\ \emph {et~al.}(2018)\citenamefont {Higashikawa}, \citenamefont {Fujita},\ and\ \citenamefont {Sato}}]{higashikawa2018}%
  \BibitemOpen
  \bibfield  {author} {\bibinfo {author} {\bibfnamefont {S.}~\bibnamefont {Higashikawa}}, \bibinfo {author} {\bibfnamefont {H.}~\bibnamefont {Fujita}},\ and\ \bibinfo {author} {\bibfnamefont {M.}~\bibnamefont {Sato}},\ }\href@noop {} {\bibinfo {title} {Floquet engineering of classical systems}} (\bibinfo {year} {2018}),\ \Eprint {https://arxiv.org/abs/1810.01103} {arXiv:1810.01103 [cond-mat.str-el]} \BibitemShut {NoStop}%
\bibitem [{\citenamefont {Mori}(2018)}]{Mori2018}%
  \BibitemOpen
  \bibfield  {author} {\bibinfo {author} {\bibfnamefont {T.}~\bibnamefont {Mori}},\ }\bibfield  {title} {\bibinfo {title} {Floquet prethermalization in periodically driven classical spin systems},\ }\href@noop {} {\bibfield  {journal} {\bibinfo  {journal} {Phys. Rev. B}\ }\textbf {\bibinfo {volume} {98}},\ \bibinfo {pages} {104303} (\bibinfo {year} {2018})}\BibitemShut {NoStop}%
\bibitem [{\citenamefont {Mori}(2022)}]{Mori2022}%
  \BibitemOpen
  \bibfield  {author} {\bibinfo {author} {\bibfnamefont {T.}~\bibnamefont {Mori}},\ }\bibfield  {title} {\bibinfo {title} {Heating rates under fast periodic driving beyond linear response},\ }\href@noop {} {\bibfield  {journal} {\bibinfo  {journal} {Phys. Rev. Lett.}\ }\textbf {\bibinfo {volume} {128}},\ \bibinfo {pages} {050604} (\bibinfo {year} {2022})}\BibitemShut {NoStop}%
\bibitem [{\citenamefont {Sun}\ and\ \citenamefont {Eckardt}(2020)}]{Sun2020}%
  \BibitemOpen
  \bibfield  {author} {\bibinfo {author} {\bibfnamefont {G.}~\bibnamefont {Sun}}\ and\ \bibinfo {author} {\bibfnamefont {A.}~\bibnamefont {Eckardt}},\ }\bibfield  {title} {\bibinfo {title} {Optimal frequency window for floquet engineering in optical lattices},\ }\href@noop {} {\bibfield  {journal} {\bibinfo  {journal} {Phys. Rev. Res.}\ }\textbf {\bibinfo {volume} {2}},\ \bibinfo {pages} {013241} (\bibinfo {year} {2020})}\BibitemShut {NoStop}%
\bibitem [{\citenamefont {Penin}\ and\ \citenamefont {Su}(2024)}]{Penin2024}%
  \BibitemOpen
  \bibfield  {author} {\bibinfo {author} {\bibfnamefont {A.~A.}\ \bibnamefont {Penin}}\ and\ \bibinfo {author} {\bibfnamefont {A.}~\bibnamefont {Su}},\ }\bibfield  {title} {\bibinfo {title} {Effective theory of classical and quantum particle dynamics in rapidly oscillating fields},\ }\href@noop {} {\bibfield  {journal} {\bibinfo  {journal} {Phys. Rev. Lett.}\ }\textbf {\bibinfo {volume} {132}},\ \bibinfo {pages} {051601} (\bibinfo {year} {2024})}\BibitemShut {NoStop}%
\bibitem [{\citenamefont {Werschnik}\ and\ \citenamefont {Gross}(2007)}]{Werschnik2007}%
  \BibitemOpen
  \bibfield  {author} {\bibinfo {author} {\bibfnamefont {J.}~\bibnamefont {Werschnik}}\ and\ \bibinfo {author} {\bibfnamefont {E.~K.~U.}\ \bibnamefont {Gross}},\ }\bibfield  {title} {\bibinfo {title} {Quantum optimal control theory},\ }\href {https://doi.org/10.1088/0953-4075/40/18/R01} {\bibfield  {journal} {\bibinfo  {journal} {Journal of Physics B: Atomic, Molecular and Optical Physics}\ }\textbf {\bibinfo {volume} {40}},\ \bibinfo {pages} {R175} (\bibinfo {year} {2007})}\BibitemShut {NoStop}%
\bibitem [{\citenamefont {Bukov}(2018)}]{Bukov2018}%
  \BibitemOpen
  \bibfield  {author} {\bibinfo {author} {\bibfnamefont {M.}~\bibnamefont {Bukov}},\ }\bibfield  {title} {\bibinfo {title} {Reinforcement learning for autonomous preparation of floquet-engineered states: Inverting the quantum kapitza oscillator},\ }\href@noop {} {\bibfield  {journal} {\bibinfo  {journal} {Phys. Rev. B}\ }\textbf {\bibinfo {volume} {98}},\ \bibinfo {pages} {224305} (\bibinfo {year} {2018})}\BibitemShut {NoStop}%
\bibitem [{\citenamefont {Castro}\ \emph {et~al.}(2022)\citenamefont {Castro}, \citenamefont {De~Giovannini}, \citenamefont {Sato}, \citenamefont {H\"ubener},\ and\ \citenamefont {Rubio}}]{Castro2022}%
  \BibitemOpen
  \bibfield  {author} {\bibinfo {author} {\bibfnamefont {A.}~\bibnamefont {Castro}}, \bibinfo {author} {\bibfnamefont {U.}~\bibnamefont {De~Giovannini}}, \bibinfo {author} {\bibfnamefont {S.~A.}\ \bibnamefont {Sato}}, \bibinfo {author} {\bibfnamefont {H.}~\bibnamefont {H\"ubener}},\ and\ \bibinfo {author} {\bibfnamefont {A.}~\bibnamefont {Rubio}},\ }\bibfield  {title} {\bibinfo {title} {Floquet engineering the band structure of materials with optimal control theory},\ }\href@noop {} {\bibfield  {journal} {\bibinfo  {journal} {Phys. Rev. Res.}\ }\textbf {\bibinfo {volume} {4}},\ \bibinfo {pages} {033213} (\bibinfo {year} {2022})}\BibitemShut {NoStop}%
\bibitem [{\citenamefont {Matsos}\ \emph {et~al.}(2024{\natexlab{a}})\citenamefont {Matsos}, \citenamefont {Valahu}, \citenamefont {Navickas}, \citenamefont {Rao}, \citenamefont {Millican}, \citenamefont {Kolesnikow}, \citenamefont {Biercuk},\ and\ \citenamefont {Tan}}]{Matsos2024}%
  \BibitemOpen
  \bibfield  {author} {\bibinfo {author} {\bibfnamefont {V.~G.}\ \bibnamefont {Matsos}}, \bibinfo {author} {\bibfnamefont {C.~H.}\ \bibnamefont {Valahu}}, \bibinfo {author} {\bibfnamefont {T.}~\bibnamefont {Navickas}}, \bibinfo {author} {\bibfnamefont {A.~D.}\ \bibnamefont {Rao}}, \bibinfo {author} {\bibfnamefont {M.~J.}\ \bibnamefont {Millican}}, \bibinfo {author} {\bibfnamefont {X.~C.}\ \bibnamefont {Kolesnikow}}, \bibinfo {author} {\bibfnamefont {M.~J.}\ \bibnamefont {Biercuk}},\ and\ \bibinfo {author} {\bibfnamefont {T.~R.}\ \bibnamefont {Tan}},\ }\bibfield  {title} {\bibinfo {title} {Robust and deterministic preparation of bosonic logical states in a trapped ion},\ }\href@noop {} {\bibfield  {journal} {\bibinfo  {journal} {Phys. Rev. Lett.}\ }\textbf {\bibinfo {volume} {133}},\ \bibinfo {pages} {050602} (\bibinfo {year} {2024}{\natexlab{a}})}\BibitemShut {NoStop}%
\bibitem [{\citenamefont {Matsos}\ \emph {et~al.}(2024{\natexlab{b}})\citenamefont {Matsos}, \citenamefont {Valahu}, \citenamefont {Millican}, \citenamefont {Navickas}, \citenamefont {Kolesnikow}, \citenamefont {Biercuk},\ and\ \citenamefont {Tan}}]{VGMatsos2024}%
  \BibitemOpen
  \bibfield  {author} {\bibinfo {author} {\bibfnamefont {V.~G.}\ \bibnamefont {Matsos}}, \bibinfo {author} {\bibfnamefont {C.~H.}\ \bibnamefont {Valahu}}, \bibinfo {author} {\bibfnamefont {M.~J.}\ \bibnamefont {Millican}}, \bibinfo {author} {\bibfnamefont {T.}~\bibnamefont {Navickas}}, \bibinfo {author} {\bibfnamefont {X.~C.}\ \bibnamefont {Kolesnikow}}, \bibinfo {author} {\bibfnamefont {M.~J.}\ \bibnamefont {Biercuk}},\ and\ \bibinfo {author} {\bibfnamefont {T.~R.}\ \bibnamefont {Tan}},\ }\href@noop {} {\bibinfo {title} {Universal quantum gate set for gottesman-kitaev-preskill logical qubits}} (\bibinfo {year} {2024}{\natexlab{b}}),\ \Eprint {https://arxiv.org/abs/2409.05455} {arXiv:2409.05455 [quant-ph]} \BibitemShut {NoStop}%
\bibitem [{\citenamefont {Evered}\ \emph {et~al.}(2023)\citenamefont {Evered}, \citenamefont {Bluvstein}, \citenamefont {Kalinowski}, \citenamefont {Ebadi}, \citenamefont {Manovitz}, \citenamefont {Zhou}, \citenamefont {Li}, \citenamefont {Geim}, \citenamefont {Wang}, \citenamefont {Maskara} \emph {et~al.}}]{evered2023}%
  \BibitemOpen
  \bibfield  {author} {\bibinfo {author} {\bibfnamefont {S.~J.}\ \bibnamefont {Evered}}, \bibinfo {author} {\bibfnamefont {D.}~\bibnamefont {Bluvstein}}, \bibinfo {author} {\bibfnamefont {M.}~\bibnamefont {Kalinowski}}, \bibinfo {author} {\bibfnamefont {S.}~\bibnamefont {Ebadi}}, \bibinfo {author} {\bibfnamefont {T.}~\bibnamefont {Manovitz}}, \bibinfo {author} {\bibfnamefont {H.}~\bibnamefont {Zhou}}, \bibinfo {author} {\bibfnamefont {S.~H.}\ \bibnamefont {Li}}, \bibinfo {author} {\bibfnamefont {A.~A.}\ \bibnamefont {Geim}}, \bibinfo {author} {\bibfnamefont {T.~T.}\ \bibnamefont {Wang}}, \bibinfo {author} {\bibfnamefont {N.}~\bibnamefont {Maskara}}, \emph {et~al.},\ }\bibfield  {title} {\bibinfo {title} {High-fidelity parallel entangling gates on a neutral-atom quantum computer},\ }\href@noop {} {\bibfield  {journal} {\bibinfo  {journal} {Nature}\ }\textbf {\bibinfo {volume} {622}},\ \bibinfo {pages} {268} (\bibinfo {year} {2023})}\BibitemShut {NoStop}%
\bibitem [{\citenamefont {Manetsch}\ \emph {et~al.}(2024)\citenamefont {Manetsch}, \citenamefont {Nomura}, \citenamefont {Bataille}, \citenamefont {Leung}, \citenamefont {Lv},\ and\ \citenamefont {Endres}}]{manetsch2024}%
  \BibitemOpen
  \bibfield  {author} {\bibinfo {author} {\bibfnamefont {H.~J.}\ \bibnamefont {Manetsch}}, \bibinfo {author} {\bibfnamefont {G.}~\bibnamefont {Nomura}}, \bibinfo {author} {\bibfnamefont {E.}~\bibnamefont {Bataille}}, \bibinfo {author} {\bibfnamefont {K.~H.}\ \bibnamefont {Leung}}, \bibinfo {author} {\bibfnamefont {X.}~\bibnamefont {Lv}},\ and\ \bibinfo {author} {\bibfnamefont {M.}~\bibnamefont {Endres}},\ }\bibfield  {title} {\bibinfo {title} {A tweezer array with 6100 highly coherent atomic qubits},\ }\href@noop {} {\bibfield  {journal} {\bibinfo  {journal} {arXiv preprint arXiv:2403.12021}\ } (\bibinfo {year} {2024})}\BibitemShut {NoStop}%
\bibitem [{\citenamefont {Li}\ \emph {et~al.}(2024)\citenamefont {Li}, \citenamefont {Kubo}, \citenamefont {Ho}, \citenamefont {Yan}, \citenamefont {Nakamura},\ and\ \citenamefont {Goto}}]{Li2024}%
  \BibitemOpen
  \bibfield  {author} {\bibinfo {author} {\bibfnamefont {R.}~\bibnamefont {Li}}, \bibinfo {author} {\bibfnamefont {K.}~\bibnamefont {Kubo}}, \bibinfo {author} {\bibfnamefont {Y.}~\bibnamefont {Ho}}, \bibinfo {author} {\bibfnamefont {Z.}~\bibnamefont {Yan}}, \bibinfo {author} {\bibfnamefont {Y.}~\bibnamefont {Nakamura}},\ and\ \bibinfo {author} {\bibfnamefont {H.}~\bibnamefont {Goto}},\ }\bibfield  {title} {\bibinfo {title} {Realization of high-fidelity cz gate based on a double-transmon coupler},\ }\href@noop {} {\bibfield  {journal} {\bibinfo  {journal} {Phys. Rev. X}\ }\textbf {\bibinfo {volume} {14}},\ \bibinfo {pages} {041050} (\bibinfo {year} {2024})}\BibitemShut {NoStop}%
\bibitem [{\citenamefont {L{\"o}schnauer}\ \emph {et~al.}(2024)\citenamefont {L{\"o}schnauer}, \citenamefont {Toba}, \citenamefont {Hughes}, \citenamefont {King}, \citenamefont {Weber}, \citenamefont {Srinivas}, \citenamefont {Matt}, \citenamefont {Nourshargh}, \citenamefont {Allcock}, \citenamefont {Ballance} \emph {et~al.}}]{loschnauer2024}%
  \BibitemOpen
  \bibfield  {author} {\bibinfo {author} {\bibfnamefont {C.}~\bibnamefont {L{\"o}schnauer}}, \bibinfo {author} {\bibfnamefont {J.~M.}\ \bibnamefont {Toba}}, \bibinfo {author} {\bibfnamefont {A.}~\bibnamefont {Hughes}}, \bibinfo {author} {\bibfnamefont {S.}~\bibnamefont {King}}, \bibinfo {author} {\bibfnamefont {M.}~\bibnamefont {Weber}}, \bibinfo {author} {\bibfnamefont {R.}~\bibnamefont {Srinivas}}, \bibinfo {author} {\bibfnamefont {R.}~\bibnamefont {Matt}}, \bibinfo {author} {\bibfnamefont {R.}~\bibnamefont {Nourshargh}}, \bibinfo {author} {\bibfnamefont {D.}~\bibnamefont {Allcock}}, \bibinfo {author} {\bibfnamefont {C.}~\bibnamefont {Ballance}}, \emph {et~al.},\ }\bibfield  {title} {\bibinfo {title} {Scalable, high-fidelity all-electronic control of trapped-ion qubits},\ }\href@noop {} {\bibfield  {journal} {\bibinfo  {journal} {arXiv preprint arXiv:2407.07694}\ } (\bibinfo {year} {2024})}\BibitemShut {NoStop}%
\bibitem [{\citenamefont {Junge}\ \emph {et~al.}(2021)\citenamefont {Junge}, \citenamefont {Frey}, \citenamefont {Ashcroft},\ and\ \citenamefont {Kügeler}}]{Junge2021}%
  \BibitemOpen
  \bibfield  {author} {\bibinfo {author} {\bibfnamefont {L.}~\bibnamefont {Junge}}, \bibinfo {author} {\bibfnamefont {C.}~\bibnamefont {Frey}}, \bibinfo {author} {\bibfnamefont {G.}~\bibnamefont {Ashcroft}},\ and\ \bibinfo {author} {\bibfnamefont {E.}~\bibnamefont {Kügeler}},\ }\bibfield  {title} {\bibinfo {title} {A new harmonic balance approach using multidimensional time},\ }\href@noop {} {\bibfield  {journal} {\bibinfo  {journal} {Journal of Engineering for Gas Turbines and Power}\ }\textbf {\bibinfo {volume} {143}},\ \bibinfo {pages} {081007} (\bibinfo {year} {2021})}\BibitemShut {NoStop}%
\bibitem [{\citenamefont {Rockafellar}(1993)}]{Rockafellar1993}%
  \BibitemOpen
  \bibfield  {author} {\bibinfo {author} {\bibfnamefont {R.~T.}\ \bibnamefont {Rockafellar}},\ }\bibfield  {title} {\bibinfo {title} {Lagrange multipliers and optimality},\ }\href@noop {} {\bibfield  {journal} {\bibinfo  {journal} {SIAM Review}\ }\textbf {\bibinfo {volume} {35}},\ \bibinfo {pages} {183} (\bibinfo {year} {1993})}\BibitemShut {NoStop}%
\bibitem [{\citenamefont {Ablowitz}\ and\ \citenamefont {Segur}(1981)}]{ablowitz1981}%
  \BibitemOpen
  \bibfield  {author} {\bibinfo {author} {\bibfnamefont {M.~J.}\ \bibnamefont {Ablowitz}}\ and\ \bibinfo {author} {\bibfnamefont {H.}~\bibnamefont {Segur}},\ }\bibinfo {title} {Solitons and the inverse scattering transform}\ (\bibinfo  {publisher} {SIAM},\ \bibinfo {year} {1981})\BibitemShut {NoStop}%
\bibitem [{\citenamefont {Kielpinski}\ \emph {et~al.}(2002)\citenamefont {Kielpinski}, \citenamefont {Monroe},\ and\ \citenamefont {Wineland}}]{kielpinski2002}%
  \BibitemOpen
  \bibfield  {author} {\bibinfo {author} {\bibfnamefont {D.}~\bibnamefont {Kielpinski}}, \bibinfo {author} {\bibfnamefont {C.}~\bibnamefont {Monroe}},\ and\ \bibinfo {author} {\bibfnamefont {D.~J.}\ \bibnamefont {Wineland}},\ }\bibfield  {title} {\bibinfo {title} {Architecture for a large-scale ion-trap quantum computer},\ }\href@noop {} {\bibfield  {journal} {\bibinfo  {journal} {Nature}\ }\textbf {\bibinfo {volume} {417}},\ \bibinfo {pages} {709} (\bibinfo {year} {2002})}\BibitemShut {NoStop}%
\bibitem [{\citenamefont {Leibfried}\ \emph {et~al.}(2003)\citenamefont {Leibfried}, \citenamefont {Blatt}, \citenamefont {Monroe},\ and\ \citenamefont {Wineland}}]{Leibfried2003}%
  \BibitemOpen
  \bibfield  {author} {\bibinfo {author} {\bibfnamefont {D.}~\bibnamefont {Leibfried}}, \bibinfo {author} {\bibfnamefont {R.}~\bibnamefont {Blatt}}, \bibinfo {author} {\bibfnamefont {C.}~\bibnamefont {Monroe}},\ and\ \bibinfo {author} {\bibfnamefont {D.}~\bibnamefont {Wineland}},\ }\bibfield  {title} {\bibinfo {title} {Quantum dynamics of single trapped ions},\ }\href@noop {} {\bibfield  {journal} {\bibinfo  {journal} {Rev. Mod. Phys.}\ }\textbf {\bibinfo {volume} {75}},\ \bibinfo {pages} {281} (\bibinfo {year} {2003})}\BibitemShut {NoStop}%
\bibitem [{\citenamefont {Häffner}\ \emph {et~al.}(2008)\citenamefont {Häffner}, \citenamefont {Roos},\ and\ \citenamefont {Blatt}}]{HAFFNER2008}%
  \BibitemOpen
  \bibfield  {author} {\bibinfo {author} {\bibfnamefont {H.}~\bibnamefont {Häffner}}, \bibinfo {author} {\bibfnamefont {C.}~\bibnamefont {Roos}},\ and\ \bibinfo {author} {\bibfnamefont {R.}~\bibnamefont {Blatt}},\ }\bibfield  {title} {\bibinfo {title} {Quantum computing with trapped ions},\ }\href@noop {} {\bibfield  {journal} {\bibinfo  {journal} {Physics Reports}\ }\textbf {\bibinfo {volume} {469}},\ \bibinfo {pages} {155} (\bibinfo {year} {2008})}\BibitemShut {NoStop}%
\bibitem [{\citenamefont {Li}\ \emph {et~al.}(2012)\citenamefont {Li}, \citenamefont {Gong}, \citenamefont {Yin}, \citenamefont {Quan}, \citenamefont {Yin}, \citenamefont {Zhang}, \citenamefont {Duan},\ and\ \citenamefont {Zhang}}]{Li2012}%
  \BibitemOpen
  \bibfield  {author} {\bibinfo {author} {\bibfnamefont {T.}~\bibnamefont {Li}}, \bibinfo {author} {\bibfnamefont {Z.-X.}\ \bibnamefont {Gong}}, \bibinfo {author} {\bibfnamefont {Z.-Q.}\ \bibnamefont {Yin}}, \bibinfo {author} {\bibfnamefont {H.~T.}\ \bibnamefont {Quan}}, \bibinfo {author} {\bibfnamefont {X.}~\bibnamefont {Yin}}, \bibinfo {author} {\bibfnamefont {P.}~\bibnamefont {Zhang}}, \bibinfo {author} {\bibfnamefont {L.-M.}\ \bibnamefont {Duan}},\ and\ \bibinfo {author} {\bibfnamefont {X.}~\bibnamefont {Zhang}},\ }\bibfield  {title} {\bibinfo {title} {Space-time crystals of trapped ions},\ }\href@noop {} {\bibfield  {journal} {\bibinfo  {journal} {Phys. Rev. Lett.}\ }\textbf {\bibinfo {volume} {109}},\ \bibinfo {pages} {163001} (\bibinfo {year} {2012})}\BibitemShut {NoStop}%
\bibitem [{\citenamefont {Yu}\ \emph {et~al.}(2022)\citenamefont {Yu}, \citenamefont {Alonso}, \citenamefont {Caminiti}, \citenamefont {Beck}, \citenamefont {Sutherland}, \citenamefont {Leibfried}, \citenamefont {Rodriguez}, \citenamefont {Dhital}, \citenamefont {Hemmerling},\ and\ \citenamefont {H\"affner}}]{Qian2022}%
  \BibitemOpen
  \bibfield  {author} {\bibinfo {author} {\bibfnamefont {Q.}~\bibnamefont {Yu}}, \bibinfo {author} {\bibfnamefont {A.~M.}\ \bibnamefont {Alonso}}, \bibinfo {author} {\bibfnamefont {J.}~\bibnamefont {Caminiti}}, \bibinfo {author} {\bibfnamefont {K.~M.}\ \bibnamefont {Beck}}, \bibinfo {author} {\bibfnamefont {R.~T.}\ \bibnamefont {Sutherland}}, \bibinfo {author} {\bibfnamefont {D.}~\bibnamefont {Leibfried}}, \bibinfo {author} {\bibfnamefont {K.~J.}\ \bibnamefont {Rodriguez}}, \bibinfo {author} {\bibfnamefont {M.}~\bibnamefont {Dhital}}, \bibinfo {author} {\bibfnamefont {B.}~\bibnamefont {Hemmerling}},\ and\ \bibinfo {author} {\bibfnamefont {H.}~\bibnamefont {H\"affner}},\ }\bibfield  {title} {\bibinfo {title} {Feasibility study of quantum computing using trapped electrons},\ }\href@noop {} {\bibfield  {journal} {\bibinfo  {journal} {Phys. Rev. A}\ }\textbf {\bibinfo {volume} {105}},\ \bibinfo {pages} {022420} (\bibinfo {year} {2022})}\BibitemShut {NoStop}%
\bibitem [{\citenamefont {Xu}\ \emph {et~al.}(2023)\citenamefont {Xu}, \citenamefont {Xia}, \citenamefont {Yu}, \citenamefont {Khan}, \citenamefont {Megidish}, \citenamefont {You}, \citenamefont {Hemmerling}, \citenamefont {Jayich}, \citenamefont {Biener},\ and\ \citenamefont {H{\"a}ffner}}]{xu2023}%
  \BibitemOpen
  \bibfield  {author} {\bibinfo {author} {\bibfnamefont {S.}~\bibnamefont {Xu}}, \bibinfo {author} {\bibfnamefont {X.}~\bibnamefont {Xia}}, \bibinfo {author} {\bibfnamefont {Q.}~\bibnamefont {Yu}}, \bibinfo {author} {\bibfnamefont {S.}~\bibnamefont {Khan}}, \bibinfo {author} {\bibfnamefont {E.}~\bibnamefont {Megidish}}, \bibinfo {author} {\bibfnamefont {B.}~\bibnamefont {You}}, \bibinfo {author} {\bibfnamefont {B.}~\bibnamefont {Hemmerling}}, \bibinfo {author} {\bibfnamefont {A.}~\bibnamefont {Jayich}}, \bibinfo {author} {\bibfnamefont {J.}~\bibnamefont {Biener}},\ and\ \bibinfo {author} {\bibfnamefont {H.}~\bibnamefont {H{\"a}ffner}},\ }\bibfield  {title} {\bibinfo {title} {3d-printed micro ion trap technology for scalable quantum information processing},\ }\href@noop {} {\bibfield  {journal} {\bibinfo  {journal} {arXiv preprint arXiv:2310.00595}\ } (\bibinfo {year} {2023})}\BibitemShut {NoStop}%
\bibitem [{\citenamefont {Liu}\ \emph {et~al.}(2023)\citenamefont {Liu}, \citenamefont {Du}, \citenamefont {Huang}, \citenamefont {He}, \citenamefont {He}, \citenamefont {Zhang}, \citenamefont {Tang}, \citenamefont {Meng}, \citenamefont {Zhai}, \citenamefont {Han},\ and\ \citenamefont {Xie}}]{Liu2023}%
  \BibitemOpen
  \bibfield  {author} {\bibinfo {author} {\bibfnamefont {Y.}~\bibnamefont {Liu}}, \bibinfo {author} {\bibfnamefont {L.}~\bibnamefont {Du}}, \bibinfo {author} {\bibfnamefont {S.}~\bibnamefont {Huang}}, \bibinfo {author} {\bibfnamefont {Y.}~\bibnamefont {He}}, \bibinfo {author} {\bibfnamefont {K.}~\bibnamefont {He}}, \bibinfo {author} {\bibfnamefont {Q.}~\bibnamefont {Zhang}}, \bibinfo {author} {\bibfnamefont {Y.}~\bibnamefont {Tang}}, \bibinfo {author} {\bibfnamefont {Y.}~\bibnamefont {Meng}}, \bibinfo {author} {\bibfnamefont {S.}~\bibnamefont {Zhai}}, \bibinfo {author} {\bibfnamefont {H.}~\bibnamefont {Han}},\ and\ \bibinfo {author} {\bibfnamefont {J.}~\bibnamefont {Xie}},\ }\bibfield  {title} {\bibinfo {title} {Quantitative assessment and suppression of anharmonic potential of quadrupole linear radiofrequency ion traps with round electrodes},\ }\href {https://doi.org/https://doi.org/10.1016/j.ijms.2022.116997} {\bibfield  {journal} {\bibinfo  {journal} {International Journal of Mass Spectrometry}\ }\textbf
  {\bibinfo {volume} {485}},\ \bibinfo {pages} {116997} (\bibinfo {year} {2023})}\BibitemShut {NoStop}%
\bibitem [{\citenamefont {Huang}\ \emph {et~al.}(2025)\citenamefont {Huang}, \citenamefont {Hausten}, \citenamefont {Yu}, \citenamefont {Taniguchi}, \citenamefont {Yadav}, \citenamefont {Sacksteder}, \citenamefont {Noguchi}, \citenamefont {Schneider},\ and\ \citenamefont {Haeffner}}]{Andris2025}%
  \BibitemOpen
  \bibfield  {author} {\bibinfo {author} {\bibfnamefont {A.}~\bibnamefont {Huang}}, \bibinfo {author} {\bibfnamefont {E.}~\bibnamefont {Hausten}}, \bibinfo {author} {\bibfnamefont {Q.}~\bibnamefont {Yu}}, \bibinfo {author} {\bibfnamefont {K.}~\bibnamefont {Taniguchi}}, \bibinfo {author} {\bibfnamefont {N.}~\bibnamefont {Yadav}}, \bibinfo {author} {\bibfnamefont {I.}~\bibnamefont {Sacksteder}}, \bibinfo {author} {\bibfnamefont {A.}~\bibnamefont {Noguchi}}, \bibinfo {author} {\bibfnamefont {R.}~\bibnamefont {Schneider}},\ and\ \bibinfo {author} {\bibfnamefont {H.}~\bibnamefont {Haeffner}},\ }\href {https://arxiv.org/abs/2503.12379} {\bibinfo {title} {Numerical investigations of electron dynamics in a linear paul trap}} (\bibinfo {year} {2025}),\ \Eprint {https://arxiv.org/abs/2503.12379} {arXiv:2503.12379 [quant-ph]} \BibitemShut {NoStop}%
\bibitem [{\citenamefont {Cameron}\ and\ \citenamefont {Griffin}(1989)}]{Cameron1989}%
  \BibitemOpen
  \bibfield  {author} {\bibinfo {author} {\bibfnamefont {T.~M.}\ \bibnamefont {Cameron}}\ and\ \bibinfo {author} {\bibfnamefont {J.~H.}\ \bibnamefont {Griffin}},\ }\bibfield  {title} {\bibinfo {title} {An alternating frequency/time domain method for calculating the steady-state response of nonlinear dynamic systems},\ }\href@noop {} {\bibfield  {journal} {\bibinfo  {journal} {Journal of Applied Mechanics}\ }\textbf {\bibinfo {volume} {56}},\ \bibinfo {pages} {149} (\bibinfo {year} {1989})}\BibitemShut {NoStop}%
\bibitem [{\citenamefont {Yan}\ \emph {et~al.}(2023)\citenamefont {Yan}, \citenamefont {Dai}, \citenamefont {Wang},\ and\ \citenamefont {Atluri}}]{Zipu2023}%
  \BibitemOpen
  \bibfield  {author} {\bibinfo {author} {\bibfnamefont {Z.}~\bibnamefont {Yan}}, \bibinfo {author} {\bibfnamefont {H.}~\bibnamefont {Dai}}, \bibinfo {author} {\bibfnamefont {Q.}~\bibnamefont {Wang}},\ and\ \bibinfo {author} {\bibfnamefont {S.~N.}\ \bibnamefont {Atluri}},\ }\bibfield  {title} {\bibinfo {title} {Harmonic balance methods: A review and recent developments},\ }\href@noop {} {\bibfield  {journal} {\bibinfo  {journal} {Computer Modeling in Engineering \& Sciences}\ }\textbf {\bibinfo {volume} {137}},\ \bibinfo {pages} {1419} (\bibinfo {year} {2023})}\BibitemShut {NoStop}%
\bibitem [{\citenamefont {Nicks}\ \emph {et~al.}(2024)\citenamefont {Nicks}, \citenamefont {Allen},\ and\ \citenamefont {Coombes}}]{Nicks2024}%
  \BibitemOpen
  \bibfield  {author} {\bibinfo {author} {\bibfnamefont {R.}~\bibnamefont {Nicks}}, \bibinfo {author} {\bibfnamefont {R.}~\bibnamefont {Allen}},\ and\ \bibinfo {author} {\bibfnamefont {S.}~\bibnamefont {Coombes}},\ }\bibfield  {title} {\bibinfo {title} {Phase and amplitude responses for delay equations using harmonic balance},\ }\href@noop {} {\bibfield  {journal} {\bibinfo  {journal} {Phys. Rev. E}\ }\textbf {\bibinfo {volume} {110}},\ \bibinfo {pages} {L012202} (\bibinfo {year} {2024})}\BibitemShut {NoStop}%
\bibitem [{\citenamefont {Lindblad}\ \emph {et~al.}(2022)\citenamefont {Lindblad}, \citenamefont {Frey}, \citenamefont {Junge}, \citenamefont {Ashcroft},\ and\ \citenamefont {Andersson}}]{lindblad2022}%
  \BibitemOpen
  \bibfield  {author} {\bibinfo {author} {\bibfnamefont {D.}~\bibnamefont {Lindblad}}, \bibinfo {author} {\bibfnamefont {C.}~\bibnamefont {Frey}}, \bibinfo {author} {\bibfnamefont {L.}~\bibnamefont {Junge}}, \bibinfo {author} {\bibfnamefont {G.}~\bibnamefont {Ashcroft}},\ and\ \bibinfo {author} {\bibfnamefont {N.}~\bibnamefont {Andersson}},\ }\bibfield  {title} {\bibinfo {title} {Minimizing aliasing in multiple frequency harmonic balance computations},\ }\href@noop {} {\bibfield  {journal} {\bibinfo  {journal} {Journal of Scientific Computing}\ }\textbf {\bibinfo {volume} {91}},\ \bibinfo {pages} {65} (\bibinfo {year} {2022})}\BibitemShut {NoStop}%
\bibitem [{\citenamefont {Doroudi}(2009)}]{Doroudi2009}%
  \BibitemOpen
  \bibfield  {author} {\bibinfo {author} {\bibfnamefont {A.}~\bibnamefont {Doroudi}},\ }\bibfield  {title} {\bibinfo {title} {Calculation of coupled secular oscillation frequencies and axial secular frequency in a nonlinear ion trap by a homotopy method},\ }\href@noop {} {\bibfield  {journal} {\bibinfo  {journal} {Phys. Rev. E}\ }\textbf {\bibinfo {volume} {80}},\ \bibinfo {pages} {056603} (\bibinfo {year} {2009})}\BibitemShut {NoStop}%
\bibitem [{\citenamefont {Goldman}\ and\ \citenamefont {Gabrielse}(2010)}]{Goldman2010}%
  \BibitemOpen
  \bibfield  {author} {\bibinfo {author} {\bibfnamefont {J.}~\bibnamefont {Goldman}}\ and\ \bibinfo {author} {\bibfnamefont {G.}~\bibnamefont {Gabrielse}},\ }\bibfield  {title} {\bibinfo {title} {Optimized planar penning traps for quantum-information studies},\ }\href@noop {} {\bibfield  {journal} {\bibinfo  {journal} {Phys. Rev. A}\ }\textbf {\bibinfo {volume} {81}},\ \bibinfo {pages} {052335} (\bibinfo {year} {2010})}\BibitemShut {NoStop}%
\bibitem [{\citenamefont {Tarantola}(2005)}]{Tarantola2005}%
  \BibitemOpen
  \bibfield  {author} {\bibinfo {author} {\bibfnamefont {A.}~\bibnamefont {Tarantola}},\ }\bibinfo {title} {Inverse problem theory and methods for model parameter estimation}\ (\bibinfo  {publisher} {Society for Industrial and Applied Mathematics},\ \bibinfo {address} {Philadelphia},\ \bibinfo {year} {2005})\BibitemShut {NoStop}%
\bibitem [{\citenamefont {Peng}\ \emph {et~al.}(2017)\citenamefont {Peng}, \citenamefont {Matthiesen},\ and\ \citenamefont {H\"affner}}]{Peng2017}%
  \BibitemOpen
  \bibfield  {author} {\bibinfo {author} {\bibfnamefont {P.}~\bibnamefont {Peng}}, \bibinfo {author} {\bibfnamefont {C.}~\bibnamefont {Matthiesen}},\ and\ \bibinfo {author} {\bibfnamefont {H.}~\bibnamefont {H\"affner}},\ }\bibfield  {title} {\bibinfo {title} {Spin readout of trapped electron qubits},\ }\href@noop {} {\bibfield  {journal} {\bibinfo  {journal} {Phys. Rev. A}\ }\textbf {\bibinfo {volume} {95}},\ \bibinfo {pages} {012312} (\bibinfo {year} {2017})}\BibitemShut {NoStop}%
\bibitem [{\citenamefont {Campbell}\ and\ \citenamefont {Hamilton}(2017)}]{campbell2017}%
  \BibitemOpen
  \bibfield  {author} {\bibinfo {author} {\bibfnamefont {W.}~\bibnamefont {Campbell}}\ and\ \bibinfo {author} {\bibfnamefont {P.}~\bibnamefont {Hamilton}},\ }\bibfield  {title} {\bibinfo {title} {Rotation sensing with trapped ions},\ }\href@noop {} {\bibfield  {journal} {\bibinfo  {journal} {Journal of Physics B: Atomic, Molecular and Optical Physics}\ }\textbf {\bibinfo {volume} {50}},\ \bibinfo {pages} {064002} (\bibinfo {year} {2017})}\BibitemShut {NoStop}%
\bibitem [{\citenamefont {Shinjo}\ \emph {et~al.}(2021)\citenamefont {Shinjo}, \citenamefont {Baba}, \citenamefont {Higashiyama}, \citenamefont {Saito},\ and\ \citenamefont {Mukaiyama}}]{Shinjo2021}%
  \BibitemOpen
  \bibfield  {author} {\bibinfo {author} {\bibfnamefont {A.}~\bibnamefont {Shinjo}}, \bibinfo {author} {\bibfnamefont {M.}~\bibnamefont {Baba}}, \bibinfo {author} {\bibfnamefont {K.}~\bibnamefont {Higashiyama}}, \bibinfo {author} {\bibfnamefont {R.}~\bibnamefont {Saito}},\ and\ \bibinfo {author} {\bibfnamefont {T.}~\bibnamefont {Mukaiyama}},\ }\bibfield  {title} {\bibinfo {title} {Three-dimensional matter-wave interferometry of a trapped single ion},\ }\href@noop {} {\bibfield  {journal} {\bibinfo  {journal} {Phys. Rev. Lett.}\ }\textbf {\bibinfo {volume} {126}},\ \bibinfo {pages} {153604} (\bibinfo {year} {2021})}\BibitemShut {NoStop}%
\bibitem [{\citenamefont {Guggemos}\ \emph {et~al.}(2015)\citenamefont {Guggemos}, \citenamefont {Heinrich}, \citenamefont {Herrera-Sancho}, \citenamefont {Blatt},\ and\ \citenamefont {Roos}}]{Guggemos2015}%
  \BibitemOpen
  \bibfield  {author} {\bibinfo {author} {\bibfnamefont {M.}~\bibnamefont {Guggemos}}, \bibinfo {author} {\bibfnamefont {D.}~\bibnamefont {Heinrich}}, \bibinfo {author} {\bibfnamefont {O.~A.}\ \bibnamefont {Herrera-Sancho}}, \bibinfo {author} {\bibfnamefont {R.}~\bibnamefont {Blatt}},\ and\ \bibinfo {author} {\bibfnamefont {C.~F.}\ \bibnamefont {Roos}},\ }\bibfield  {title} {\bibinfo {title} {Sympathetic cooling and detection of a hot trapped ion by a cold one},\ }\href {https://doi.org/10.1088/1367-2630/17/10/103001} {\bibfield  {journal} {\bibinfo  {journal} {New Journal of Physics}\ }\textbf {\bibinfo {volume} {17}},\ \bibinfo {pages} {103001} (\bibinfo {year} {2015})}\BibitemShut {NoStop}%
\bibitem [{\citenamefont {Jauffred}\ \emph {et~al.}(2017)\citenamefont {Jauffred}, \citenamefont {Onofrio},\ and\ \citenamefont {Sundaram}}]{Francisco2017}%
  \BibitemOpen
  \bibfield  {author} {\bibinfo {author} {\bibfnamefont {F.}~\bibnamefont {Jauffred}}, \bibinfo {author} {\bibfnamefont {R.}~\bibnamefont {Onofrio}},\ and\ \bibinfo {author} {\bibfnamefont {B.}~\bibnamefont {Sundaram}},\ }\bibfield  {title} {\bibinfo {title} {Simulating sympathetic cooling of atomic mixtures in nonlinear traps},\ }\href@noop {} {\bibfield  {journal} {\bibinfo  {journal} {Physics Letters A}\ }\textbf {\bibinfo {volume} {381}},\ \bibinfo {pages} {2783} (\bibinfo {year} {2017})}\BibitemShut {NoStop}%
\bibitem [{\citenamefont {Bohman}\ \emph {et~al.}(2021)\citenamefont {Bohman}, \citenamefont {Grunhofer}, \citenamefont {Smorra}, \citenamefont {Wiesinger}, \citenamefont {Will}, \citenamefont {Borchert}, \citenamefont {Devlin}, \citenamefont {Erlewein}, \citenamefont {Fleck}, \citenamefont {Gavranovic}, \citenamefont {Harrington}, \citenamefont {Latacz}, \citenamefont {Mooser}, \citenamefont {Popper}, \citenamefont {Wursten}, \citenamefont {Blaum}, \citenamefont {Matsuda}, \citenamefont {Ospelkaus}, \citenamefont {Quint}, \citenamefont {Walz}, \citenamefont {Ulmer},\ and\ \citenamefont {Collaboration}}]{Bohman2021}%
  \BibitemOpen
  \bibfield  {author} {\bibinfo {author} {\bibfnamefont {M.}~\bibnamefont {Bohman}}, \bibinfo {author} {\bibfnamefont {V.}~\bibnamefont {Grunhofer}}, \bibinfo {author} {\bibfnamefont {C.}~\bibnamefont {Smorra}}, \bibinfo {author} {\bibfnamefont {M.}~\bibnamefont {Wiesinger}}, \bibinfo {author} {\bibfnamefont {C.}~\bibnamefont {Will}}, \bibinfo {author} {\bibfnamefont {M.~J.}\ \bibnamefont {Borchert}}, \bibinfo {author} {\bibfnamefont {J.~A.}\ \bibnamefont {Devlin}}, \bibinfo {author} {\bibfnamefont {S.}~\bibnamefont {Erlewein}}, \bibinfo {author} {\bibfnamefont {M.}~\bibnamefont {Fleck}}, \bibinfo {author} {\bibfnamefont {S.}~\bibnamefont {Gavranovic}}, \bibinfo {author} {\bibfnamefont {J.}~\bibnamefont {Harrington}}, \bibinfo {author} {\bibfnamefont {B.}~\bibnamefont {Latacz}}, \bibinfo {author} {\bibfnamefont {A.}~\bibnamefont {Mooser}}, \bibinfo {author} {\bibfnamefont {D.}~\bibnamefont {Popper}}, \bibinfo {author} {\bibfnamefont {E.}~\bibnamefont {Wursten}}, \bibinfo {author} {\bibfnamefont
  {K.}~\bibnamefont {Blaum}}, \bibinfo {author} {\bibfnamefont {Y.}~\bibnamefont {Matsuda}}, \bibinfo {author} {\bibfnamefont {C.}~\bibnamefont {Ospelkaus}}, \bibinfo {author} {\bibfnamefont {W.}~\bibnamefont {Quint}}, \bibinfo {author} {\bibfnamefont {J.}~\bibnamefont {Walz}}, \bibinfo {author} {\bibfnamefont {S.}~\bibnamefont {Ulmer}},\ and\ \bibinfo {author} {\bibfnamefont {B.~A. S.~E.}\ \bibnamefont {Collaboration}},\ }\bibfield  {title} {\bibinfo {title} {Sympathetic cooling of a trapped proton mediated by an lc circuit},\ }\href@noop {} {\bibfield  {journal} {\bibinfo  {journal} {Nature}\ }\textbf {\bibinfo {volume} {596}},\ \bibinfo {pages} {514} (\bibinfo {year} {2021})}\BibitemShut {NoStop}%
\bibitem [{\citenamefont {Home}\ \emph {et~al.}(2011)\citenamefont {Home}, \citenamefont {Hanneke}, \citenamefont {Jost}, \citenamefont {Leibfried},\ and\ \citenamefont {Wineland}}]{Home2011}%
  \BibitemOpen
  \bibfield  {author} {\bibinfo {author} {\bibfnamefont {J.~P.}\ \bibnamefont {Home}}, \bibinfo {author} {\bibfnamefont {D.}~\bibnamefont {Hanneke}}, \bibinfo {author} {\bibfnamefont {J.~D.}\ \bibnamefont {Jost}}, \bibinfo {author} {\bibfnamefont {D.}~\bibnamefont {Leibfried}},\ and\ \bibinfo {author} {\bibfnamefont {D.~J.}\ \bibnamefont {Wineland}},\ }\bibfield  {title} {\bibinfo {title} {Normal modes of trapped ions in the presence of anharmonic trap potentials},\ }\href@noop {} {\bibfield  {journal} {\bibinfo  {journal} {New Journal of Physics}\ }\textbf {\bibinfo {volume} {13}},\ \bibinfo {pages} {073026} (\bibinfo {year} {2011})}\BibitemShut {NoStop}%
\bibitem [{\citenamefont {Sutherland}\ \emph {et~al.}(2022)\citenamefont {Sutherland}, \citenamefont {Yu}, \citenamefont {Beck},\ and\ \citenamefont {H\"affner}}]{Sutherland2022}%
  \BibitemOpen
  \bibfield  {author} {\bibinfo {author} {\bibfnamefont {R.~T.}\ \bibnamefont {Sutherland}}, \bibinfo {author} {\bibfnamefont {Q.}~\bibnamefont {Yu}}, \bibinfo {author} {\bibfnamefont {K.~M.}\ \bibnamefont {Beck}},\ and\ \bibinfo {author} {\bibfnamefont {H.}~\bibnamefont {H\"affner}},\ }\bibfield  {title} {\bibinfo {title} {One- and two-qubit gate infidelities due to motional errors in trapped ions and electrons},\ }\href@noop {} {\bibfield  {journal} {\bibinfo  {journal} {Phys. Rev. A}\ }\textbf {\bibinfo {volume} {105}},\ \bibinfo {pages} {022437} (\bibinfo {year} {2022})}\BibitemShut {NoStop}%
\bibitem [{\citenamefont {Gottesman}\ \emph {et~al.}(2001)\citenamefont {Gottesman}, \citenamefont {Kitaev},\ and\ \citenamefont {Preskill}}]{Gottesman2001}%
  \BibitemOpen
  \bibfield  {author} {\bibinfo {author} {\bibfnamefont {D.}~\bibnamefont {Gottesman}}, \bibinfo {author} {\bibfnamefont {A.}~\bibnamefont {Kitaev}},\ and\ \bibinfo {author} {\bibfnamefont {J.}~\bibnamefont {Preskill}},\ }\bibfield  {title} {\bibinfo {title} {Encoding a qubit in an oscillator},\ }\href@noop {} {\bibfield  {journal} {\bibinfo  {journal} {Phys. Rev. A}\ }\textbf {\bibinfo {volume} {64}},\ \bibinfo {pages} {012310} (\bibinfo {year} {2001})}\BibitemShut {NoStop}%
\bibitem [{\citenamefont {Fl{\"u}hmann}\ \emph {et~al.}(2019{\natexlab{a}})\citenamefont {Fl{\"u}hmann}, \citenamefont {Nguyen}, \citenamefont {Marinelli}, \citenamefont {Negnevitsky}, \citenamefont {Mehta},\ and\ \citenamefont {Home}}]{Flühmann2019}%
  \BibitemOpen
  \bibfield  {author} {\bibinfo {author} {\bibfnamefont {C.}~\bibnamefont {Fl{\"u}hmann}}, \bibinfo {author} {\bibfnamefont {T.~L.}\ \bibnamefont {Nguyen}}, \bibinfo {author} {\bibfnamefont {M.}~\bibnamefont {Marinelli}}, \bibinfo {author} {\bibfnamefont {V.}~\bibnamefont {Negnevitsky}}, \bibinfo {author} {\bibfnamefont {K.}~\bibnamefont {Mehta}},\ and\ \bibinfo {author} {\bibfnamefont {J.~P.}\ \bibnamefont {Home}},\ }\bibfield  {title} {\bibinfo {title} {Encoding a qubit in a trapped-ion mechanical oscillator},\ }\href@noop {} {\bibfield  {journal} {\bibinfo  {journal} {Nature}\ }\textbf {\bibinfo {volume} {566}},\ \bibinfo {pages} {513} (\bibinfo {year} {2019}{\natexlab{a}})}\BibitemShut {NoStop}%
\bibitem [{\citenamefont {Fl{\"u}hmann}\ \emph {et~al.}(2019{\natexlab{b}})\citenamefont {Fl{\"u}hmann}, \citenamefont {Nguyen}, \citenamefont {Marinelli}, \citenamefont {Negnevitsky}, \citenamefont {Mehta},\ and\ \citenamefont {Home}}]{fluhmann2019}%
  \BibitemOpen
  \bibfield  {author} {\bibinfo {author} {\bibfnamefont {C.}~\bibnamefont {Fl{\"u}hmann}}, \bibinfo {author} {\bibfnamefont {T.~L.}\ \bibnamefont {Nguyen}}, \bibinfo {author} {\bibfnamefont {M.}~\bibnamefont {Marinelli}}, \bibinfo {author} {\bibfnamefont {V.}~\bibnamefont {Negnevitsky}}, \bibinfo {author} {\bibfnamefont {K.}~\bibnamefont {Mehta}},\ and\ \bibinfo {author} {\bibfnamefont {J.}~\bibnamefont {Home}},\ }\bibfield  {title} {\bibinfo {title} {Encoding a qubit in a trapped-ion mechanical oscillator},\ }\href@noop {} {\bibfield  {journal} {\bibinfo  {journal} {Nature}\ }\textbf {\bibinfo {volume} {566}},\ \bibinfo {pages} {513} (\bibinfo {year} {2019}{\natexlab{b}})}\BibitemShut {NoStop}%
\bibitem [{\citenamefont {Rojkov}\ \emph {et~al.}(2024)\citenamefont {Rojkov}, \citenamefont {R\"oggla}, \citenamefont {Wagener}, \citenamefont {Fontbot\'e-Schmidt}, \citenamefont {Welte}, \citenamefont {Home},\ and\ \citenamefont {Reiter}}]{Rojkov2024}%
  \BibitemOpen
  \bibfield  {author} {\bibinfo {author} {\bibfnamefont {I.}~\bibnamefont {Rojkov}}, \bibinfo {author} {\bibfnamefont {P.~M.}\ \bibnamefont {R\"oggla}}, \bibinfo {author} {\bibfnamefont {M.}~\bibnamefont {Wagener}}, \bibinfo {author} {\bibfnamefont {M.}~\bibnamefont {Fontbot\'e-Schmidt}}, \bibinfo {author} {\bibfnamefont {S.}~\bibnamefont {Welte}}, \bibinfo {author} {\bibfnamefont {J.}~\bibnamefont {Home}},\ and\ \bibinfo {author} {\bibfnamefont {F.}~\bibnamefont {Reiter}},\ }\bibfield  {title} {\bibinfo {title} {Two-qubit operations for finite-energy gottesman-kitaev-preskill encodings},\ }\href@noop {} {\bibfield  {journal} {\bibinfo  {journal} {Phys. Rev. Lett.}\ }\textbf {\bibinfo {volume} {133}},\ \bibinfo {pages} {100601} (\bibinfo {year} {2024})}\BibitemShut {NoStop}%
\bibitem [{\citenamefont {Bohnmann}\ \emph {et~al.}(2024)\citenamefont {Bohnmann}, \citenamefont {Locher}, \citenamefont {Zeiher},\ and\ \citenamefont {Müller}}]{Leon2024}%
  \BibitemOpen
  \bibfield  {author} {\bibinfo {author} {\bibfnamefont {L.~H.}\ \bibnamefont {Bohnmann}}, \bibinfo {author} {\bibfnamefont {D.~F.}\ \bibnamefont {Locher}}, \bibinfo {author} {\bibfnamefont {J.}~\bibnamefont {Zeiher}},\ and\ \bibinfo {author} {\bibfnamefont {M.}~\bibnamefont {Müller}},\ }\href@noop {} {\bibinfo {title} {Bosonic quantum error correction with neutral atoms in optical dipole traps}} (\bibinfo {year} {2024}),\ \Eprint {https://arxiv.org/abs/2408.14251} {arXiv:2408.14251 [quant-ph]} \BibitemShut {NoStop}%
\bibitem [{\citenamefont {Schmied}\ \emph {et~al.}(2009)\citenamefont {Schmied}, \citenamefont {Wesenberg},\ and\ \citenamefont {Leibfried}}]{Schmied2009}%
  \BibitemOpen
  \bibfield  {author} {\bibinfo {author} {\bibfnamefont {R.}~\bibnamefont {Schmied}}, \bibinfo {author} {\bibfnamefont {J.~H.}\ \bibnamefont {Wesenberg}},\ and\ \bibinfo {author} {\bibfnamefont {D.}~\bibnamefont {Leibfried}},\ }\bibfield  {title} {\bibinfo {title} {Optimal surface-electrode trap lattices for quantum simulation with trapped ions},\ }\href@noop {} {\bibfield  {journal} {\bibinfo  {journal} {Phys. Rev. Lett.}\ }\textbf {\bibinfo {volume} {102}},\ \bibinfo {pages} {233002} (\bibinfo {year} {2009})}\BibitemShut {NoStop}%
\bibitem [{\citenamefont {Kim}\ \emph {et~al.}(2010)\citenamefont {Kim}, \citenamefont {Herskind}, \citenamefont {Kim}, \citenamefont {Kim},\ and\ \citenamefont {Chuang}}]{Kim2010}%
  \BibitemOpen
  \bibfield  {author} {\bibinfo {author} {\bibfnamefont {T.~H.}\ \bibnamefont {Kim}}, \bibinfo {author} {\bibfnamefont {P.~F.}\ \bibnamefont {Herskind}}, \bibinfo {author} {\bibfnamefont {T.}~\bibnamefont {Kim}}, \bibinfo {author} {\bibfnamefont {J.}~\bibnamefont {Kim}},\ and\ \bibinfo {author} {\bibfnamefont {I.~L.}\ \bibnamefont {Chuang}},\ }\bibfield  {title} {\bibinfo {title} {Surface-electrode point paul trap},\ }\href {https://doi.org/10.1103/PhysRevA.82.043412} {\bibfield  {journal} {\bibinfo  {journal} {Phys. Rev. A}\ }\textbf {\bibinfo {volume} {82}},\ \bibinfo {pages} {043412} (\bibinfo {year} {2010})}\BibitemShut {NoStop}%
\bibitem [{\citenamefont {Romaszko}\ \emph {et~al.}(2020)\citenamefont {Romaszko}, \citenamefont {Hong}, \citenamefont {Siegele}, \citenamefont {Puddy}, \citenamefont {Lebrun-Gallagher}, \citenamefont {Weidt},\ and\ \citenamefont {Hensinger}}]{romaszko2020}%
  \BibitemOpen
  \bibfield  {author} {\bibinfo {author} {\bibfnamefont {Z.~D.}\ \bibnamefont {Romaszko}}, \bibinfo {author} {\bibfnamefont {S.}~\bibnamefont {Hong}}, \bibinfo {author} {\bibfnamefont {M.}~\bibnamefont {Siegele}}, \bibinfo {author} {\bibfnamefont {R.~K.}\ \bibnamefont {Puddy}}, \bibinfo {author} {\bibfnamefont {F.~R.}\ \bibnamefont {Lebrun-Gallagher}}, \bibinfo {author} {\bibfnamefont {S.}~\bibnamefont {Weidt}},\ and\ \bibinfo {author} {\bibfnamefont {W.~K.}\ \bibnamefont {Hensinger}},\ }\bibfield  {title} {\bibinfo {title} {Engineering of microfabricated ion traps and integration of advanced on-chip features},\ }\href@noop {} {\bibfield  {journal} {\bibinfo  {journal} {Nature Reviews Physics}\ }\textbf {\bibinfo {volume} {2}},\ \bibinfo {pages} {285} (\bibinfo {year} {2020})}\BibitemShut {NoStop}%
\bibitem [{\citenamefont {Palani}\ \emph {et~al.}(2023)\citenamefont {Palani}, \citenamefont {Hasse}, \citenamefont {Kiefer}, \citenamefont {Boeckling}, \citenamefont {Schroeder}, \citenamefont {Warring},\ and\ \citenamefont {Schaetz}}]{Palani2023}%
  \BibitemOpen
  \bibfield  {author} {\bibinfo {author} {\bibfnamefont {D.}~\bibnamefont {Palani}}, \bibinfo {author} {\bibfnamefont {F.}~\bibnamefont {Hasse}}, \bibinfo {author} {\bibfnamefont {P.}~\bibnamefont {Kiefer}}, \bibinfo {author} {\bibfnamefont {F.}~\bibnamefont {Boeckling}}, \bibinfo {author} {\bibfnamefont {J.-P.}\ \bibnamefont {Schroeder}}, \bibinfo {author} {\bibfnamefont {U.}~\bibnamefont {Warring}},\ and\ \bibinfo {author} {\bibfnamefont {T.}~\bibnamefont {Schaetz}},\ }\bibfield  {title} {\bibinfo {title} {High-fidelity transport of trapped-ion qubits in a multilayer array},\ }\href@noop {} {\bibfield  {journal} {\bibinfo  {journal} {Phys. Rev. A}\ }\textbf {\bibinfo {volume} {107}},\ \bibinfo {pages} {L050601} (\bibinfo {year} {2023})}\BibitemShut {NoStop}%
\bibitem [{\citenamefont {Seidling}\ \emph {et~al.}(2024)\citenamefont {Seidling}, \citenamefont {Schmidt-Kaler}, \citenamefont {Zimmermann}, \citenamefont {Simonaitis}, \citenamefont {Keathley}, \citenamefont {Berggren},\ and\ \citenamefont {Hommelhoff}}]{Seidling2024}%
  \BibitemOpen
  \bibfield  {author} {\bibinfo {author} {\bibfnamefont {M.}~\bibnamefont {Seidling}}, \bibinfo {author} {\bibfnamefont {F.~D.~F.}\ \bibnamefont {Schmidt-Kaler}}, \bibinfo {author} {\bibfnamefont {R.}~\bibnamefont {Zimmermann}}, \bibinfo {author} {\bibfnamefont {J.~W.}\ \bibnamefont {Simonaitis}}, \bibinfo {author} {\bibfnamefont {P.~D.}\ \bibnamefont {Keathley}}, \bibinfo {author} {\bibfnamefont {K.~K.}\ \bibnamefont {Berggren}},\ and\ \bibinfo {author} {\bibfnamefont {P.}~\bibnamefont {Hommelhoff}},\ }\bibfield  {title} {\bibinfo {title} {Resonating electrostatically guided electrons},\ }\href@noop {} {\bibfield  {journal} {\bibinfo  {journal} {Phys. Rev. Lett.}\ }\textbf {\bibinfo {volume} {132}},\ \bibinfo {pages} {255001} (\bibinfo {year} {2024})}\BibitemShut {NoStop}%
\bibitem [{\citenamefont {Lysne}\ \emph {et~al.}(2024)\citenamefont {Lysne}, \citenamefont {Niedermeyer}, \citenamefont {Wilson}, \citenamefont {Slichter},\ and\ \citenamefont {Leibfried}}]{Lysne2024}%
  \BibitemOpen
  \bibfield  {author} {\bibinfo {author} {\bibfnamefont {N.~K.}\ \bibnamefont {Lysne}}, \bibinfo {author} {\bibfnamefont {J.~F.}\ \bibnamefont {Niedermeyer}}, \bibinfo {author} {\bibfnamefont {A.~C.}\ \bibnamefont {Wilson}}, \bibinfo {author} {\bibfnamefont {D.~H.}\ \bibnamefont {Slichter}},\ and\ \bibinfo {author} {\bibfnamefont {D.}~\bibnamefont {Leibfried}},\ }\bibfield  {title} {\bibinfo {title} {Individual addressing and state readout of trapped ions utilizing radio-frequency micromotion},\ }\href@noop {} {\bibfield  {journal} {\bibinfo  {journal} {Phys. Rev. Lett.}\ }\textbf {\bibinfo {volume} {133}},\ \bibinfo {pages} {033201} (\bibinfo {year} {2024})}\BibitemShut {NoStop}%
\bibitem [{\citenamefont {Pietik\"ainen}\ \emph {et~al.}(2019)\citenamefont {Pietik\"ainen}, \citenamefont {Tuorila}, \citenamefont {Golubev},\ and\ \citenamefont {Paraoanu}}]{Pietik2019}%
  \BibitemOpen
  \bibfield  {author} {\bibinfo {author} {\bibfnamefont {I.}~\bibnamefont {Pietik\"ainen}}, \bibinfo {author} {\bibfnamefont {J.}~\bibnamefont {Tuorila}}, \bibinfo {author} {\bibfnamefont {D.~S.}\ \bibnamefont {Golubev}},\ and\ \bibinfo {author} {\bibfnamefont {G.~S.}\ \bibnamefont {Paraoanu}},\ }\bibfield  {title} {\bibinfo {title} {Photon blockade and the quantum-to-classical transition in the driven-dissipative josephson pendulum coupled to a resonator},\ }\href {https://doi.org/10.1103/PhysRevA.99.063828} {\bibfield  {journal} {\bibinfo  {journal} {Phys. Rev. A}\ }\textbf {\bibinfo {volume} {99}},\ \bibinfo {pages} {063828} (\bibinfo {year} {2019})}\BibitemShut {NoStop}%
\bibitem [{\citenamefont {Frattini}\ \emph {et~al.}(2018)\citenamefont {Frattini}, \citenamefont {Sivak}, \citenamefont {Lingenfelter}, \citenamefont {Shankar},\ and\ \citenamefont {Devoret}}]{Frattini2018}%
  \BibitemOpen
  \bibfield  {author} {\bibinfo {author} {\bibfnamefont {N.~E.}\ \bibnamefont {Frattini}}, \bibinfo {author} {\bibfnamefont {V.~V.}\ \bibnamefont {Sivak}}, \bibinfo {author} {\bibfnamefont {A.}~\bibnamefont {Lingenfelter}}, \bibinfo {author} {\bibfnamefont {S.}~\bibnamefont {Shankar}},\ and\ \bibinfo {author} {\bibfnamefont {M.~H.}\ \bibnamefont {Devoret}},\ }\bibfield  {title} {\bibinfo {title} {Optimizing the nonlinearity and dissipation of a snail parametric amplifier for dynamic range},\ }\href {https://doi.org/10.1103/PhysRevApplied.10.054020} {\bibfield  {journal} {\bibinfo  {journal} {Phys. Rev. Appl.}\ }\textbf {\bibinfo {volume} {10}},\ \bibinfo {pages} {054020} (\bibinfo {year} {2018})}\BibitemShut {NoStop}%
\bibitem [{\citenamefont {Bahar}\ and\ \citenamefont {Ba{\c{s}}er}(2022)}]{bahar2022}%
  \BibitemOpen
  \bibfield  {author} {\bibinfo {author} {\bibfnamefont {M.}~\bibnamefont {Bahar}}\ and\ \bibinfo {author} {\bibfnamefont {P.}~\bibnamefont {Ba{\c{s}}er}},\ }\bibfield  {title} {\bibinfo {title} {Tuning of nonlinear optical characteristics of mathieu quantum dot by laser and electric field},\ }\href@noop {} {\bibfield  {journal} {\bibinfo  {journal} {The European Physical Journal Plus}\ }\textbf {\bibinfo {volume} {137}},\ \bibinfo {pages} {1138} (\bibinfo {year} {2022})}\BibitemShut {NoStop}%
\bibitem [{\citenamefont {Blakestad}\ \emph {et~al.}(2009)\citenamefont {Blakestad}, \citenamefont {Ospelkaus}, \citenamefont {VanDevender}, \citenamefont {Amini}, \citenamefont {Britton}, \citenamefont {Leibfried},\ and\ \citenamefont {Wineland}}]{blakestad2009}%
  \BibitemOpen
  \bibfield  {author} {\bibinfo {author} {\bibfnamefont {R.}~\bibnamefont {Blakestad}}, \bibinfo {author} {\bibfnamefont {C.}~\bibnamefont {Ospelkaus}}, \bibinfo {author} {\bibfnamefont {A.}~\bibnamefont {VanDevender}}, \bibinfo {author} {\bibfnamefont {J.}~\bibnamefont {Amini}}, \bibinfo {author} {\bibfnamefont {J.}~\bibnamefont {Britton}}, \bibinfo {author} {\bibfnamefont {D.}~\bibnamefont {Leibfried}},\ and\ \bibinfo {author} {\bibfnamefont {D.~J.}\ \bibnamefont {Wineland}},\ }\bibfield  {title} {\bibinfo {title} {High-fidelity transport of trapped-ion qubits through an x-junction trap array},\ }\href@noop {} {\bibfield  {journal} {\bibinfo  {journal} {Physical review letters}\ }\textbf {\bibinfo {volume} {102}},\ \bibinfo {pages} {153002} (\bibinfo {year} {2009})}\BibitemShut {NoStop}%
\bibitem [{\citenamefont {Sterk}\ \emph {et~al.}(2022)\citenamefont {Sterk}, \citenamefont {Coakley}, \citenamefont {Goldberg}, \citenamefont {Hietala}, \citenamefont {Lechtenberg}, \citenamefont {McGuinness}, \citenamefont {McMurtrey}, \citenamefont {Parazzoli}, \citenamefont {Van Der~Wall},\ and\ \citenamefont {Stick}}]{sterk2022}%
  \BibitemOpen
  \bibfield  {author} {\bibinfo {author} {\bibfnamefont {J.~D.}\ \bibnamefont {Sterk}}, \bibinfo {author} {\bibfnamefont {H.}~\bibnamefont {Coakley}}, \bibinfo {author} {\bibfnamefont {J.}~\bibnamefont {Goldberg}}, \bibinfo {author} {\bibfnamefont {V.}~\bibnamefont {Hietala}}, \bibinfo {author} {\bibfnamefont {J.}~\bibnamefont {Lechtenberg}}, \bibinfo {author} {\bibfnamefont {H.}~\bibnamefont {McGuinness}}, \bibinfo {author} {\bibfnamefont {D.}~\bibnamefont {McMurtrey}}, \bibinfo {author} {\bibfnamefont {L.~P.}\ \bibnamefont {Parazzoli}}, \bibinfo {author} {\bibfnamefont {J.}~\bibnamefont {Van Der~Wall}},\ and\ \bibinfo {author} {\bibfnamefont {D.}~\bibnamefont {Stick}},\ }\bibfield  {title} {\bibinfo {title} {Closed-loop optimization of fast trapped-ion shuttling with sub-quanta excitation},\ }\href@noop {} {\bibfield  {journal} {\bibinfo  {journal} {npj Quantum Information}\ }\textbf {\bibinfo {volume} {8}},\ \bibinfo {pages} {68} (\bibinfo {year} {2022})}\BibitemShut {NoStop}%
\bibitem [{\citenamefont {Delaney}\ \emph {et~al.}(2024)\citenamefont {Delaney}, \citenamefont {Sletten}, \citenamefont {Cich}, \citenamefont {Estey}, \citenamefont {Fabrikant}, \citenamefont {Hayes}, \citenamefont {Hoffman}, \citenamefont {Hostetter}, \citenamefont {Langer}, \citenamefont {Moses} \emph {et~al.}}]{delaney2024}%
  \BibitemOpen
  \bibfield  {author} {\bibinfo {author} {\bibfnamefont {R.~D.}\ \bibnamefont {Delaney}}, \bibinfo {author} {\bibfnamefont {L.~R.}\ \bibnamefont {Sletten}}, \bibinfo {author} {\bibfnamefont {M.~J.}\ \bibnamefont {Cich}}, \bibinfo {author} {\bibfnamefont {B.}~\bibnamefont {Estey}}, \bibinfo {author} {\bibfnamefont {M.~I.}\ \bibnamefont {Fabrikant}}, \bibinfo {author} {\bibfnamefont {D.}~\bibnamefont {Hayes}}, \bibinfo {author} {\bibfnamefont {I.~M.}\ \bibnamefont {Hoffman}}, \bibinfo {author} {\bibfnamefont {J.}~\bibnamefont {Hostetter}}, \bibinfo {author} {\bibfnamefont {C.}~\bibnamefont {Langer}}, \bibinfo {author} {\bibfnamefont {S.~A.}\ \bibnamefont {Moses}}, \emph {et~al.},\ }\bibfield  {title} {\bibinfo {title} {Scalable multispecies ion transport in a grid-based surface-electrode trap},\ }\href@noop {} {\bibfield  {journal} {\bibinfo  {journal} {Physical Review X}\ }\textbf {\bibinfo {volume} {14}},\ \bibinfo {pages} {041028} (\bibinfo {year} {2024})}\BibitemShut {NoStop}%
\bibitem [{\citenamefont {Didier}\ \emph {et~al.}(2012)\citenamefont {Didier}, \citenamefont {Sinou},\ and\ \citenamefont {Faverjon}}]{Didier2012}%
  \BibitemOpen
  \bibfield  {author} {\bibinfo {author} {\bibfnamefont {J.}~\bibnamefont {Didier}}, \bibinfo {author} {\bibfnamefont {J.-J.}\ \bibnamefont {Sinou}},\ and\ \bibinfo {author} {\bibfnamefont {B.}~\bibnamefont {Faverjon}},\ }\bibfield  {title} {\bibinfo {title} {Multi-dimensional harmonic balance with uncertainties applied to rotor dynamics},\ }\href@noop {} {\bibfield  {journal} {\bibinfo  {journal} {Journal of Vibration and Acoustics}\ }\textbf {\bibinfo {volume} {134}},\ \bibinfo {pages} {061003} (\bibinfo {year} {2012})}\BibitemShut {NoStop}%
\bibitem [{\citenamefont {Didier}\ \emph {et~al.}(2013)\citenamefont {Didier}, \citenamefont {Sinou},\ and\ \citenamefont {Faverjon}}]{Didier2013}%
  \BibitemOpen
  \bibfield  {author} {\bibinfo {author} {\bibfnamefont {J.}~\bibnamefont {Didier}}, \bibinfo {author} {\bibfnamefont {J.-J.}\ \bibnamefont {Sinou}},\ and\ \bibinfo {author} {\bibfnamefont {B.}~\bibnamefont {Faverjon}},\ }\bibfield  {title} {\bibinfo {title} {Nonlinear vibrations of a mechanical system with non-regular nonlinearities and uncertainties},\ }\href@noop {} {\bibfield  {journal} {\bibinfo  {journal} {Communications in Nonlinear Science and Numerical Simulation}\ }\textbf {\bibinfo {volume} {18}},\ \bibinfo {pages} {3250} (\bibinfo {year} {2013})}\BibitemShut {NoStop}%
\bibitem [{\citenamefont {Miguel}\ \emph {et~al.}(2022)\citenamefont {Miguel}, \citenamefont {Teloli},\ and\ \citenamefont {Silva}}]{Miguel2022}%
  \BibitemOpen
  \bibfield  {author} {\bibinfo {author} {\bibfnamefont {L.~P.}\ \bibnamefont {Miguel}}, \bibinfo {author} {\bibfnamefont {R.~d.~O.}\ \bibnamefont {Teloli}},\ and\ \bibinfo {author} {\bibfnamefont {S.~d.}\ \bibnamefont {Silva}},\ }\bibfield  {title} {\bibinfo {title} {Bayesian model identification through harmonic balance method for hysteresis prediction in bolted joints},\ }\href@noop {} {\bibfield  {journal} {\bibinfo  {journal} {Nonlinear Dynamics}\ }\textbf {\bibinfo {volume} {107}},\ \bibinfo {pages} {77} (\bibinfo {year} {2022})}\BibitemShut {NoStop}%
\bibitem [{\citenamefont {Fu}\ \emph {et~al.}(2023)\citenamefont {Fu}, \citenamefont {Sinou}, \citenamefont {Zhu}, \citenamefont {Lu},\ and\ \citenamefont {Yang}}]{Chao2023}%
  \BibitemOpen
  \bibfield  {author} {\bibinfo {author} {\bibfnamefont {C.}~\bibnamefont {Fu}}, \bibinfo {author} {\bibfnamefont {J.-J.}\ \bibnamefont {Sinou}}, \bibinfo {author} {\bibfnamefont {W.}~\bibnamefont {Zhu}}, \bibinfo {author} {\bibfnamefont {K.}~\bibnamefont {Lu}},\ and\ \bibinfo {author} {\bibfnamefont {Y.}~\bibnamefont {Yang}},\ }\bibfield  {title} {\bibinfo {title} {A state-of-the-art review on uncertainty analysis of rotor systems},\ }\href@noop {} {\bibfield  {journal} {\bibinfo  {journal} {Mechanical Systems and Signal Processing}\ }\textbf {\bibinfo {volume} {183}},\ \bibinfo {pages} {109619} (\bibinfo {year} {2023})}\BibitemShut {NoStop}%
\bibitem [{\citenamefont {Goldman}\ \emph {et~al.}(2023)\citenamefont {Goldman}, \citenamefont {Diessel}, \citenamefont {Barbiero}, \citenamefont {Pr\"ufer}, \citenamefont {Di~Liberto},\ and\ \citenamefont {Peralta~Gavensky}}]{Goldman2023}%
  \BibitemOpen
  \bibfield  {author} {\bibinfo {author} {\bibfnamefont {N.}~\bibnamefont {Goldman}}, \bibinfo {author} {\bibfnamefont {O.}~\bibnamefont {Diessel}}, \bibinfo {author} {\bibfnamefont {L.}~\bibnamefont {Barbiero}}, \bibinfo {author} {\bibfnamefont {M.}~\bibnamefont {Pr\"ufer}}, \bibinfo {author} {\bibfnamefont {M.}~\bibnamefont {Di~Liberto}},\ and\ \bibinfo {author} {\bibfnamefont {L.}~\bibnamefont {Peralta~Gavensky}},\ }\bibfield  {title} {\bibinfo {title} {Floquet-engineered nonlinearities and controllable pair-hopping processes: From optical kerr cavities to correlated quantum matter},\ }\href {https://doi.org/10.1103/PRXQuantum.4.040327} {\bibfield  {journal} {\bibinfo  {journal} {PRX Quantum}\ }\textbf {\bibinfo {volume} {4}},\ \bibinfo {pages} {040327} (\bibinfo {year} {2023})}\BibitemShut {NoStop}%
\bibitem [{\citenamefont {Gesla}\ \emph {et~al.}(2024)\citenamefont {Gesla}, \citenamefont {Duguet}, \citenamefont {Quéré},\ and\ \citenamefont {Witkowski}}]{gesla2024}%
  \BibitemOpen
  \bibfield  {author} {\bibinfo {author} {\bibfnamefont {A.}~\bibnamefont {Gesla}}, \bibinfo {author} {\bibfnamefont {Y.}~\bibnamefont {Duguet}}, \bibinfo {author} {\bibfnamefont {P.~L.}\ \bibnamefont {Quéré}},\ and\ \bibinfo {author} {\bibfnamefont {L.~M.}\ \bibnamefont {Witkowski}},\ }\href@noop {} {\bibinfo {title} {Stability analysis of periodic orbits in nonlinear dynamical systems using chebyshev polynomials}} (\bibinfo {year} {2024}),\ \Eprint {https://arxiv.org/abs/2407.18230} {arXiv:2407.18230 [physics.flu-dyn]} \BibitemShut {NoStop}%
\bibitem [{\citenamefont {Xu}\ and\ \citenamefont {Guo}(2025)}]{Xu2025}%
  \BibitemOpen
  \bibfield  {author} {\bibinfo {author} {\bibfnamefont {Y.}~\bibnamefont {Xu}}\ and\ \bibinfo {author} {\bibfnamefont {L.}~\bibnamefont {Guo}},\ }\bibfield  {title} {\bibinfo {title} {Perturbative framework for engineering arbitrary floquet hamiltonian},\ }\href {https://doi.org/10.1088/1361-6633/adb072} {\bibfield  {journal} {\bibinfo  {journal} {Reports on Progress in Physics}\ }\textbf {\bibinfo {volume} {88}},\ \bibinfo {pages} {037602} (\bibinfo {year} {2025})}\BibitemShut {NoStop}%
\bibitem [{\citenamefont {Siegel}\ \emph {et~al.}(2024)\citenamefont {Siegel}, \citenamefont {McGrew}, \citenamefont {Hassan}, \citenamefont {Chen}, \citenamefont {Beloy}, \citenamefont {Grogan}, \citenamefont {Zhang},\ and\ \citenamefont {Ludlow}}]{Siegel2024}%
  \BibitemOpen
  \bibfield  {author} {\bibinfo {author} {\bibfnamefont {J.~L.}\ \bibnamefont {Siegel}}, \bibinfo {author} {\bibfnamefont {W.~F.}\ \bibnamefont {McGrew}}, \bibinfo {author} {\bibfnamefont {Y.~S.}\ \bibnamefont {Hassan}}, \bibinfo {author} {\bibfnamefont {C.-C.}\ \bibnamefont {Chen}}, \bibinfo {author} {\bibfnamefont {K.}~\bibnamefont {Beloy}}, \bibinfo {author} {\bibfnamefont {T.}~\bibnamefont {Grogan}}, \bibinfo {author} {\bibfnamefont {X.}~\bibnamefont {Zhang}},\ and\ \bibinfo {author} {\bibfnamefont {A.~D.}\ \bibnamefont {Ludlow}},\ }\bibfield  {title} {\bibinfo {title} {Excited-band coherent delocalization for improved optical lattice clock performance},\ }\href@noop {} {\bibfield  {journal} {\bibinfo  {journal} {Phys. Rev. Lett.}\ }\textbf {\bibinfo {volume} {132}},\ \bibinfo {pages} {133201} (\bibinfo {year} {2024})}\BibitemShut {NoStop}%
\bibitem [{\citenamefont {Collin}\ \emph {et~al.}(2010)\citenamefont {Collin}, \citenamefont {Larson},\ and\ \citenamefont {Martikainen}}]{Collin2010}%
  \BibitemOpen
  \bibfield  {author} {\bibinfo {author} {\bibfnamefont {A.}~\bibnamefont {Collin}}, \bibinfo {author} {\bibfnamefont {J.}~\bibnamefont {Larson}},\ and\ \bibinfo {author} {\bibfnamefont {J.~P.}\ \bibnamefont {Martikainen}},\ }\bibfield  {title} {\bibinfo {title} {Quantum states of $p$-band bosons in optical lattices},\ }\href@noop {} {\bibfield  {journal} {\bibinfo  {journal} {Phys. Rev. A}\ }\textbf {\bibinfo {volume} {81}},\ \bibinfo {pages} {023605} (\bibinfo {year} {2010})}\BibitemShut {NoStop}%
\bibitem [{\citenamefont {Pinheiro}\ \emph {et~al.}(2012)\citenamefont {Pinheiro}, \citenamefont {Martikainen},\ and\ \citenamefont {Larson}}]{Pinheiro2012}%
  \BibitemOpen
  \bibfield  {author} {\bibinfo {author} {\bibfnamefont {F.}~\bibnamefont {Pinheiro}}, \bibinfo {author} {\bibfnamefont {J.-P.}\ \bibnamefont {Martikainen}},\ and\ \bibinfo {author} {\bibfnamefont {J.}~\bibnamefont {Larson}},\ }\bibfield  {title} {\bibinfo {title} {Confined $p$-band bose-einstein condensates},\ }\href@noop {} {\bibfield  {journal} {\bibinfo  {journal} {Phys. Rev. A}\ }\textbf {\bibinfo {volume} {85}},\ \bibinfo {pages} {033638} (\bibinfo {year} {2012})}\BibitemShut {NoStop}%
\bibitem [{\citenamefont {Pietraszewicz}\ \emph {et~al.}(2013)\citenamefont {Pietraszewicz}, \citenamefont {Sowi\ifmmode~\acute{n}\else \'{n}\fi{}ski}, \citenamefont {Brewczyk}, \citenamefont {Lewenstein},\ and\ \citenamefont {Gajda}}]{Pietraszewicz2013}%
  \BibitemOpen
  \bibfield  {author} {\bibinfo {author} {\bibfnamefont {J.}~\bibnamefont {Pietraszewicz}}, \bibinfo {author} {\bibfnamefont {T.}~\bibnamefont {Sowi\ifmmode~\acute{n}\else \'{n}\fi{}ski}}, \bibinfo {author} {\bibfnamefont {M.}~\bibnamefont {Brewczyk}}, \bibinfo {author} {\bibfnamefont {M.}~\bibnamefont {Lewenstein}},\ and\ \bibinfo {author} {\bibfnamefont {M.}~\bibnamefont {Gajda}},\ }\bibfield  {title} {\bibinfo {title} {Spin dynamics of two bosons in an optical lattice site: A role of anharmonicity and anisotropy of the trapping potential},\ }\href@noop {} {\bibfield  {journal} {\bibinfo  {journal} {Phys. Rev. A}\ }\textbf {\bibinfo {volume} {88}},\ \bibinfo {pages} {013608} (\bibinfo {year} {2013})}\BibitemShut {NoStop}%
\end{thebibliography}%
@article{bahar2022,
  title={Tuning of nonlinear optical characteristics of Mathieu quantum dot by laser and electric field},
  author={Bahar, MK and Ba{\c{s}}er, P},
  journal={The European Physical Journal Plus},
  volume={137},
  number={10},
  pages={1138},
  year={2022},
  publisher={Springer}
}

@article{Frattini2018,
  title = {Optimizing the Nonlinearity and Dissipation of a SNAIL Parametric Amplifier for Dynamic Range},
  author = {Frattini, N. E. and Sivak, V. V. and Lingenfelter, A. and Shankar, S. and Devoret, M. H.},
  journal = {Phys. Rev. Appl.},
  volume = {10},
  issue = {5},
  pages = {054020},
  numpages = {18},
  year = {2018},
  month = {11},
  publisher = {American Physical Society},
  doi = {10.1103/PhysRevApplied.10.054020},
  url = {https://link.aps.org/doi/10.1103/PhysRevApplied.10.054020}
}

@article{Pietik2019,
  title = {Photon blockade and the quantum-to-classical transition in the driven-dissipative Josephson pendulum coupled to a resonator},
  author = {Pietik\"ainen, I. and Tuorila, J. and Golubev, D. S. and Paraoanu, G. S.},
  journal = {Phys. Rev. A},
  volume = {99},
  issue = {6},
  pages = {063828},
  numpages = {17},
  year = {2019},
  month = {6},
  publisher = {American Physical Society},
  doi = {10.1103/PhysRevA.99.063828},
  url = {https://link.aps.org/doi/10.1103/PhysRevA.99.063828}
}

@article{Goldman2023,
  title = {Floquet-Engineered Nonlinearities and Controllable Pair-Hopping Processes: From Optical Kerr Cavities to Correlated Quantum Matter},
  author = {Goldman, N. and Diessel, O.K. and Barbiero, L. and Pr\"ufer, M. and Di Liberto, M. and Peralta Gavensky, L.},
  journal = {PRX Quantum},
  volume = {4},
  issue = {4},
  pages = {040327},
  numpages = {37},
  year = {2023},
  month = {11},
  publisher = {American Physical Society},
  doi = {10.1103/PRXQuantum.4.040327},
  url = {https://link.aps.org/doi/10.1103/PRXQuantum.4.040327}
}

@article{Kim2010,
  title = {Surface-electrode point Paul trap},
  author = {Kim, Tony Hyun and Herskind, Peter F. and Kim, Taehyun and Kim, Jungsang and Chuang, Isaac L.},
  journal = {Phys. Rev. A},
  volume = {82},
  issue = {4},
  pages = {043412},
  numpages = {9},
  year = {2010},
  month = {10},
  publisher = {American Physical Society},
  doi = {10.1103/PhysRevA.82.043412},
  url = {https://link.aps.org/doi/10.1103/PhysRevA.82.043412}
}

@article{Ding2023,
  title = {High-Fidelity, Frequency-Flexible Two-Qubit Fluxonium Gates with a Transmon Coupler},
  author = {Ding, Leon and Hays, Max and Sung, Youngkyu and Kannan, Bharath and An, Junyoung and Di Paolo, Agustin and Karamlou, Amir H. and Hazard, Thomas M. and Azar, Kate and Kim, David K. and Niedzielski, Bethany M. and Melville, Alexander and Schwartz, Mollie E. and Yoder, Jonilyn L. and Orlando, Terry P. and Gustavsson, Simon and Grover, Jeffrey A. and Serniak, Kyle and Oliver, William D.},
  journal = {Phys. Rev. X},
  volume = {13},
  issue = {3},
  pages = {031035},
  numpages = {24},
  year = {2023},
  month = {Sep},
  publisher = {American Physical Society}
}

@article{fluhmann2019,
  title={Encoding a qubit in a trapped-ion mechanical oscillator},
  author={Fl{\"u}hmann, Christa and Nguyen, Thanh Long and Marinelli, Matteo and Negnevitsky, Vlad and Mehta, Karan and Home, JP},
  journal={Nature},
  volume={566},
  number={7745},
  pages={513--517},
  year={2019},
  publisher={Nature Publishing Group UK London}
}

@article{Rojkov2024,
  title = {Two-Qubit Operations for Finite-Energy Gottesman-Kitaev-Preskill Encodings},
  author = {Rojkov, Ivan and R\"oggla, Paul Moser and Wagener, Martin and Fontbot\'e-Schmidt, Moritz and Welte, Stephan and Home, Jonathan and Reiter, Florentin},
  journal = {Phys. Rev. Lett.},
  volume = {133},
  issue = {10},
  pages = {100601},
  numpages = {7},
  year = {2024},
  month = {9},
  publisher = {American Physical Society}
}

@article{Seidling2024,
  title = {Resonating Electrostatically Guided Electrons},
  author = {Seidling, M. and Schmidt-Kaler, F. D. F. and Zimmermann, R. and Simonaitis, J. W. and Keathley, P. D. and Berggren, K. K. and Hommelhoff, P.},
  journal = {Phys. Rev. Lett.},
  volume = {132},
  issue = {25},
  pages = {255001},
  numpages = {5},
  year = {2024},
  month = {Jun},
  publisher = {American Physical Society}
}

@article{Cameron1989,
    author = {Cameron, T. M. and Griffin, J. H.},
    title = {An Alternating Frequency/Time Domain Method for Calculating the Steady-State Response of Nonlinear Dynamic Systems},
    journal = {Journal of Applied Mechanics},
    volume = {56},
    number = {1},
    pages = {149-154},
    year = {1989},
    month = {03}
}

@article{Didier2013,
title = {Nonlinear vibrations of a mechanical system with non-regular nonlinearities and uncertainties},
journal = {Communications in Nonlinear Science and Numerical Simulation},
volume = {18},
number = {11},
pages = {3250-3270},
year = {2013},
issn = {1007-5704},
author = {J. Didier and J.-J. Sinou and B. Faverjon}
}

@article{romaszko2020,
  title={Engineering of microfabricated ion traps and integration of advanced on-chip features},
  author={Romaszko, Zak David and Hong, Seokjun and Siegele, Martin and Puddy, Reuben Kahan and Lebrun-Gallagher, Foni Rapha{\"e}l and Weidt, Sebastian and Hensinger, Winfried Karl},
  journal={Nature Reviews Physics},
  volume={2},
  number={6},
  pages={285--299},
  year={2020},
  publisher={Nature Publishing Group UK London}
}

@article{Palani2023,
  title = {High-fidelity transport of trapped-ion qubits in a multilayer array},
  author = {Palani, Deviprasath and Hasse, Florian and Kiefer, Philip and Boeckling, Frederick and Schroeder, Jan-Philipp and Warring, Ulrich and Schaetz, Tobias},
  journal = {Phys. Rev. A},
  volume = {107},
  issue = {5},
  pages = {L050601},
  numpages = {6},
  year = {2023},
  month = {5},
  publisher = {American Physical Society}
}

@article{xu2023,
  title={3D-Printed Micro Ion Trap Technology for Scalable Quantum Information Processing},
  author={Xu, Shuqi and Xia, Xiaoxing and Yu, Qian and Khan, Sumanta and Megidish, Eli and You, Bingran and Hemmerling, Boerge and Jayich, Andrew and Biener, Juergen and H{\"a}ffner, Hartmut},
  journal={arXiv preprint arXiv:2310.00595},
  year={2023}
}

@misc{Andris2025,
      title={Numerical Investigations of Electron Dynamics in a Linear Paul Trap}, 
      author={Andris Huang and Edith Hausten and Qian Yu and Kento Taniguchi and Neha Yadav and Isabel Sacksteder and Atsushi Noguchi and Ralf Schneider and Hartmut Haeffner},
      year={2025},
      eprint={2503.12379},
      archivePrefix={arXiv},
      primaryClass={quant-ph},
      url={https://arxiv.org/abs/2503.12379}, 
}

@inbook{ablowitz1981,
  title={Solitons and the inverse scattering transform},
  author={Ablowitz, Mark J and Segur, Harvey},
  year={1981},
  publisher={SIAM}
}

@article{evered2023,
  title={High-fidelity parallel entangling gates on a neutral-atom quantum computer},
  author={Evered, Simon J and Bluvstein, Dolev and Kalinowski, Marcin and Ebadi, Sepehr and Manovitz, Tom and Zhou, Hengyun and Li, Sophie H and Geim, Alexandra A and Wang, Tout T and Maskara, Nishad and others},
  journal={Nature},
  volume={622},
  number={7982},
  pages={268--272},
  year={2023},
  publisher={Nature Publishing Group UK London}
}

@article{loschnauer2024,
  title={Scalable, high-fidelity all-electronic control of trapped-ion qubits},
  author={L{\"o}schnauer, CM and Toba, J Mosca and Hughes, AC and King, SA and Weber, MA and Srinivas, R and Matt, R and Nourshargh, R and Allcock, DTC and Ballance, CJ and others},
  journal={arXiv preprint arXiv:2407.07694},
  year={2024}
}

@article{Li2024,
  title = {Realization of High-Fidelity CZ Gate Based on a Double-Transmon Coupler},
  author = {Li, Rui and Kubo, Kentaro and Ho, Yinghao and Yan, Zhiguang and Nakamura, Yasunobu and Goto, Hayato},
  journal = {Phys. Rev. X},
  volume = {14},
  issue = {4},
  pages = {041050},
  numpages = {30},
  year = {2024},
  month = {11},
  publisher = {American Physical Society}
}

@article{manetsch2024,
  title={A tweezer array with 6100 highly coherent atomic qubits},
  author={Manetsch, Hannah J and Nomura, Gyohei and Bataille, Elie and Leung, Kon H and Lv, Xudong and Endres, Manuel},
  journal={arXiv preprint arXiv:2403.12021},
  year={2024}
}

@article{Yin2022,
  title = {Floquet Engineering Hz-Level Rabi Spectra in Shallow Optical Lattice Clock},
  author = {Yin, Mo-Juan and Lu, Xiao-Tong and Li, Ting and Xia, Jing-Jing and Wang, Tao and Zhang, Xue-Feng and Chang, Hong},
  journal = {Phys. Rev. Lett.},
  volume = {128},
  issue = {7},
  pages = {073603},
  numpages = {5},
  year = {2022},
  month = {2},
  publisher = {American Physical Society}
}

@article{Kiefer2019,
  title = {Floquet-Engineered Vibrational Dynamics in a Two-Dimensional Array of Trapped Ions},
  author = {Kiefer, Philip and Hakelberg, Frederick and Wittemer, Matthias and Berm\'udez, Alejandro and Porras, Diego and Warring, Ulrich and Schaetz, Tobias},
  journal = {Phys. Rev. Lett.},
  volume = {123},
  issue = {21},
  pages = {213605},
  numpages = {6},
  year = {2019},
  month = {11},
  publisher = {American Physical Society}
}

@article{nguyen2024,
  title={Programmable Heisenberg interactions between Floquet qubits},
  author={Nguyen, Long B and Kim, Yosep and Hashim, Akel and Goss, Noah and Marinelli, Brian and Bhandari, Bibek and Das, Debmalya and Naik, Ravi K and Kreikebaum, John Mark and Jordan, Andrew N and others},
  journal={Nature Physics},
  volume={20},
  number={2},
  pages={240--246},
  year={2024},
  publisher={Nature Publishing Group UK London}
}

@article{Choi2020,
  title = {Robust Dynamic Hamiltonian Engineering of Many-Body Spin Systems},
  author = {Choi, Joonhee and Zhou, Hengyun and Knowles, Helena S. and Landig, Renate and Choi, Soonwon and Lukin, Mikhail D.},
  journal = {Phys. Rev. X},
  volume = {10},
  issue = {3},
  pages = {031002},
  numpages = {27},
  year = {2020},
  month = {7},
  publisher = {American Physical Society}
}

@article{Winterfeldt2008,
  title = {Colloquium: Optimal control of high-harmonic generation},
  author = {Winterfeldt, Carsten and Spielmann, Christian and Gerber, Gustav},
  journal = {Rev. Mod. Phys.},
  volume = {80},
  issue = {1},
  pages = {117--140},
  numpages = {0},
  year = {2008},
  month = {1},
  publisher = {American Physical Society}
}

@article{Zaletel2023,
  title = {Colloquium: Quantum and classical discrete time crystals},
  author = {Zaletel, Michael P. and Lukin, Mikhail and Monroe, Christopher and Nayak, Chetan and Wilczek, Frank and Yao, Norman Y.},
  journal = {Rev. Mod. Phys.},
  volume = {95},
  issue = {3},
  pages = {031001},
  numpages = {34},
  year = {2023},
  month = {7},
  publisher = {American Physical Society}
}

@article{Eckardt2017,
  title = {Colloquium: Atomic quantum gases in periodically driven optical lattices},
  author = {Eckardt, Andr\'e},
  journal = {Rev. Mod. Phys.},
  volume = {89},
  issue = {1},
  pages = {011004},
  numpages = {30},
  year = {2017},
  month = {3},
  publisher = {American Physical Society}
}

@Article{Hensinger2001,
author={Hensinger, W. K. and H{\"a}ffner, H. and Browaeys, A. and Heckenberg, N. R. and Helmerson, K. and McKenzie, C. and Milburn, G. J. and Phillips, W. D. and Rolston, S. L. and Rubinsztein-Dunlop, H. and Upcroft, B.},
title={Dynamical tunnelling of ultracold atoms},
journal={Nature},
year={2001},
month={7},
day={01},
volume={412},
number={6842},
pages={52-55}
}

@article{Kolesnikow2024,
  title = {Gottesman-Kitaev-Preskill State Preparation Using Periodic Driving},
  author = {Kolesnikow, Xanda C. and Bomantara, Raditya W. and Doherty, Andrew C. and Grimsmo, Arne L.},
  journal = {Phys. Rev. Lett.},
  volume = {132},
  issue = {13},
  pages = {130605},
  numpages = {6},
  year = {2024},
  month = {3},
  publisher = {American Physical Society}
}

@article{Penin2024,
  title = {Effective Theory of Classical and Quantum Particle Dynamics in Rapidly Oscillating Fields},
  author = {Penin, Alexander A. and Su, Aneca},
  journal = {Phys. Rev. Lett.},
  volume = {132},
  issue = {5},
  pages = {051601},
  numpages = {5},
  year = {2024},
  month = {1},
  publisher = {American Physical Society}
}

@article{Guo2024,
  title = {Engineering Arbitrary Hamiltonians in Phase Space},
  author = {Guo, Lingzhen and Peano, Vittorio},
  journal = {Phys. Rev. Lett.},
  volume = {132},
  issue = {2},
  pages = {023602},
  numpages = {8},
  year = {2024},
  month = {1},
  publisher = {American Physical Society}
}

@article{Collin2010,
  title = {Quantum states of $p$-band bosons in optical lattices},
  author = {Collin, A. and Larson, J. and Martikainen, J. -P.},
  journal = {Phys. Rev. A},
  volume = {81},
  issue = {2},
  pages = {023605},
  numpages = {9},
  year = {2010},
  month = {2},
  publisher = {American Physical Society}
}

@article{Pinheiro2012,
  title = {Confined $p$-band Bose-Einstein condensates},
  author = {Pinheiro, Fernanda and Martikainen, Jani-Petri and Larson, Jonas},
  journal = {Phys. Rev. A},
  volume = {85},
  issue = {3},
  pages = {033638},
  numpages = {12},
  year = {2012},
  month = {3},
  publisher = {American Physical Society}
}

@article{Pietraszewicz2013,
  title = {Spin dynamics of two bosons in an optical lattice site: A role of anharmonicity and anisotropy of the trapping potential},
  author = {Pietraszewicz, Joanna and Sowi\ifmmode \acute{n}\else \'{n}\fi{}ski, Tomasz and Brewczyk, Miros\l{}aw and Lewenstein, Maciej and Gajda, Mariusz},
  journal = {Phys. Rev. A},
  volume = {88},
  issue = {1},
  pages = {013608},
  numpages = {12},
  year = {2013},
  month = {7},
  publisher = {American Physical Society}
}

@article{Siegel2024,
  title = {Excited-Band Coherent Delocalization for Improved Optical Lattice Clock Performance},
  author = {Siegel, J. L. and McGrew, W. F. and Hassan, Y. S. and Chen, C.-C. and Beloy, K. and Grogan, T. and Zhang, X. and Ludlow, A. D.},
  journal = {Phys. Rev. Lett.},
  volume = {132},
  issue = {13},
  pages = {133201},
  numpages = {9},
  year = {2024},
  month = {3},
  publisher = {American Physical Society}
}

@misc{higashikawa2018,
      title={Floquet engineering of classical systems}, 
      author={Sho Higashikawa and Hiroyuki Fujita and Masahiro Sato},
      year={2018},
      eprint={1810.01103},
      archivePrefix={arXiv},
      primaryClass={cond-mat.str-el}
}

@article{Mori2022,
  title = {Heating Rates under Fast Periodic Driving beyond Linear Response},
  author = {Mori, Takashi},
  journal = {Phys. Rev. Lett.},
  volume = {128},
  issue = {5},
  pages = {050604},
  numpages = {6},
  year = {2022},
  month = {2},
  publisher = {American Physical Society}
}

@article{Mori2018,
  title = {Floquet prethermalization in periodically driven classical spin systems},
  author = {Mori, Takashi},
  journal = {Phys. Rev. B},
  volume = {98},
  issue = {10},
  pages = {104303},
  numpages = {12},
  year = {2018},
  month = {9},
  publisher = {American Physical Society}
}

@article{Oka2019,
   author = "Oka, Takashi and Kitamura, Sota",
   title = "Floquet Engineering of Quantum Materials", 
   journal= "Annual Review of Condensed Matter Physics",
   year = "2019",
   volume = "10",
   number = "Volume 10, 2019",
   pages = "387-408",
   publisher = "Annual Reviews",
   issn = "1947-5462",
   type = "Journal Article",
   keywords = "Floquet topological systems",
   keywords = "ultrafast spintronics",
   keywords = "nonequilibrium quantum systems",
   keywords = "Mott insulator"
  }

@article{kovacic2018,
    author = {Kovacic, Ivana and Rand, Richard and Mohamed Sah, Si},
    title = {Mathieu's Equation and Its Generalizations: Overview of Stability Charts and Their Features},
    journal = {Applied Mechanics Reviews},
    volume = {70},
    number = {2},
    pages = {020802},
    year = {2018},
    month = {02}
}

@article{Stamper2013,
  title = {Spinor Bose gases: Symmetries, magnetism, and quantum dynamics},
  author = {Stamper-Kurn, Dan M. and Ueda, Masahito},
  journal = {Rev. Mod. Phys.},
  volume = {85},
  issue = {3},
  pages = {1191--1244},
  numpages = {0},
  year = {2013},
  month = {7},
  publisher = {American Physical Society},
  doi = {10.1103/RevModPhys.85.1191},
  url = {https://link.aps.org/doi/10.1103/RevModPhys.85.1191}
}

@ARTICLE{Gilbert2004,
  author={Gilbert, T.L.},
  journal={IEEE Transactions on Magnetics}, 
  title={A phenomenological theory of damping in ferromagnetic materials}, 
  year={2004},
  volume={40},
  number={6},
  pages={3443-3449},
  keywords={Damping;Magnetic materials;Magnetization;Energy loss;Eddy currents;Extraterrestrial measurements;Magnetic domains;Lattices;Magnetic field induced strain;Equations;Ferromagnetic damping;ferromagnetic materials;magnetic core memories;magnetic domains;magnetic losses;magnetic recording;magnetization processes}
}

@article{Martin2015,
author = {Marin Bukov, Luca D'Alessio and Anatoli Polkovnikov},
title = {Universal high-frequency behavior of periodically driven systems: from dynamical stabilization to Floquet engineering},
journal = {Advances in Physics},
volume = {64},
number = {2},
pages = {139--226},
year = {2015},
publisher = {Taylor \& Francis}
}

@article{Bukov2018,
  title = {Reinforcement learning for autonomous preparation of Floquet-engineered states: Inverting the quantum Kapitza oscillator},
  author = {Bukov, Marin},
  journal = {Phys. Rev. B},
  volume = {98},
  issue = {22},
  pages = {224305},
  numpages = {16},
  year = {2018},
  month = {12},
  publisher = {American Physical Society}
}

@article{Werschnik2007,
doi = {10.1088/0953-4075/40/18/R01},
url = {https://dx.doi.org/10.1088/0953-4075/40/18/R01},
year = {2007},
month = {9},
publisher = {},
volume = {40},
number = {18},
pages = {R175},
author = {J Werschnik and E K U Gross},
title = {Quantum optimal control theory},
journal = {Journal of Physics B: Atomic, Molecular and Optical Physics}
}

@article{Sun2020,
  title = {Optimal frequency window for Floquet engineering in optical lattices},
  author = {Sun, Gaoyong and Eckardt, Andr\'e},
  journal = {Phys. Rev. Res.},
  volume = {2},
  issue = {1},
  pages = {013241},
  numpages = {10},
  year = {2020},
  month = {3},
  publisher = {American Physical Society}
}

@article{Castro2022,
  title = {Floquet engineering the band structure of materials with optimal control theory},
  author = {Castro, Alberto and De Giovannini, Umberto and Sato, Shunsuke A. and H\"ubener, Hannes and Rubio, Angel},
  journal = {Phys. Rev. Res.},
  volume = {4},
  issue = {3},
  pages = {033213},
  numpages = {11},
  year = {2022},
  month = {9},
  publisher = {American Physical Society}
}

@article{Matsos2024,
  title = {Robust and Deterministic Preparation of Bosonic Logical States in a Trapped Ion},
  author = {Matsos, V. G. and Valahu, C. H. and Navickas, T. and Rao, A. D. and Millican, M. J. and Kolesnikow, X. C. and Biercuk, M. J. and Tan, T. R.},
  journal = {Phys. Rev. Lett.},
  volume = {133},
  issue = {5},
  pages = {050602},
  numpages = {7},
  year = {2024},
  month = {7},
  publisher = {American Physical Society}
}

@misc{VGMatsos2024,
      title={Universal Quantum Gate Set for Gottesman-Kitaev-Preskill Logical Qubits}, 
      author={V. G. Matsos and C. H. Valahu and M. J. Millican and T. Navickas and X. C. Kolesnikow and M. J. Biercuk and T. R. Tan},
      year={2024},
      eprint={2409.05455},
      archivePrefix={arXiv},
      primaryClass={quant-ph}
}

@article{Junge2021,
    author = {Junge, Laura and Frey, Christian and Ashcroft, Graham and Kügeler, Edmund},
    title = {A New Harmonic Balance Approach Using Multidimensional Time},
    journal = {Journal of Engineering for Gas Turbines and Power},
    volume = {143},
    number = {8},
    pages = {081007},
    year = {2021},
    month = {03}
}

@article{Genesio1992,
title = {Harmonic balance methods for the analysis of chaotic dynamics in nonlinear systems},
journal = {Automatica},
volume = {28},
number = {3},
pages = {531-548},
year = {1992},
issn = {0005-1098},
author = {R. Genesio and A. Tesi},
keywords = {Nonlinear systems, feedback, harmonic analysis, describing function, chaotic dynamics}
}

@article{Rockafellar1993,
author = {Rockafellar, R. Tyrrell},
title = {Lagrange Multipliers and Optimality},
journal = {SIAM Review},
volume = {35},
number = {2},
pages = {183-238},
year = {1993}
}

@article{Leibfried2003,
  title = {Quantum dynamics of single trapped ions},
  author = {Leibfried, D. and Blatt, R. and Monroe, C. and Wineland, D.},
  journal = {Rev. Mod. Phys.},
  volume = {75},
  issue = {1},
  pages = {281--324},
  numpages = {0},
  year = {2003},
  month = {3},
  publisher = {American Physical Society}
}

@article{Goldman2014,
  title = {Periodically Driven Quantum Systems: Effective Hamiltonians and Engineered Gauge Fields},
  author = {Goldman, N. and Dalibard, J.},
  journal = {Phys. Rev. X},
  volume = {4},
  issue = {3},
  pages = {031027},
  numpages = {29},
  year = {2014},
  month = {8},
  publisher = {American Physical Society}
}

@article{Nicks2024,
  title = {Phase and amplitude responses for delay equations using harmonic balance},
  author = {Nicks, R. and Allen, R. and Coombes, S.},
  journal = {Phys. Rev. E},
  volume = {110},
  issue = {1},
  pages = {L012202},
  numpages = {5},
  year = {2024},
  month = {7},
  publisher = {American Physical Society}
}

@article{Zipu2023,
    AUTHOR = {Zipu Yan and Honghua Dai and Qisi Wang and Satya N. Atluri},
    TITLE = {Harmonic Balance Methods: A Review and Recent Developments},
    JOURNAL = {Computer Modeling in Engineering \& Sciences},
    VOLUME = {137},
    YEAR = {2023},
    NUMBER = {2},
    PAGES = {1419-1459},
    ISSN = {1526-1506}
}

@article{Doroudi2009,
  title = {Calculation of coupled secular oscillation frequencies and axial secular frequency in a nonlinear ion trap by a homotopy method},
  author = {Doroudi, Alireza},
  journal = {Phys. Rev. E},
  volume = {80},
  issue = {5},
  pages = {056603},
  numpages = {7},
  year = {2009},
  month = {11},
  publisher = {American Physical Society}
}

@article{Goldman2010,
  title = {Optimized planar Penning traps for quantum-information studies},
  author = {Goldman, J. and Gabrielse, G.},
  journal = {Phys. Rev. A},
  volume = {81},
  issue = {5},
  pages = {052335},
  numpages = {25},
  year = {2010},
  month = {5},
  publisher = {American Physical Society}
}

@article{Home2011,
year = {2011},
month = {jul},
publisher = {},
volume = {13},
number = {7},
pages = {073026},
author = {J P Home and D Hanneke and J D Jost and D Leibfried and D J Wineland},
title = {Normal modes of trapped ions in the presence of anharmonic trap potentials},
journal = {New Journal of Physics}
}

@inbook{Tarantola2005,
author = {Tarantola, Albert},
title = {Inverse Problem Theory and Methods for Model Parameter Estimation},
publisher = {Society for Industrial and Applied Mathematics},
year = {2005},
address = {Philadelphia}
}

@article{Peng2017,
  title = {Spin readout of trapped electron qubits},
  author = {Peng, Pai and Matthiesen, Clemens and H\"affner, Hartmut},
  journal = {Phys. Rev. A},
  volume = {95},
  issue = {1},
  pages = {012312},
  numpages = {8},
  year = {2017},
  month = {Jan},
  publisher = {American Physical Society}
}

@article{Shinjo2021,
  title = {Three-Dimensional Matter-Wave Interferometry of a Trapped Single Ion},
  author = {Shinjo, Ami and Baba, Masato and Higashiyama, Koya and Saito, Ryoichi and Mukaiyama, Takashi},
  journal = {Phys. Rev. Lett.},
  volume = {126},
  issue = {15},
  pages = {153604},
  numpages = {5},
  year = {2021},
  month = {Apr},
  publisher = {American Physical Society}
}

@article{campbell2017,
  title={Rotation sensing with trapped ions},
  author={Campbell, WC and Hamilton, P},
  journal={Journal of Physics B: Atomic, Molecular and Optical Physics},
  volume={50},
  number={6},
  pages={064002},
  year={2017},
  publisher={IOP Publishing}
}

@article{Francisco2017,
title = {Simulating sympathetic cooling of atomic mixtures in nonlinear traps},
journal = {Physics Letters A},
volume = {381},
number = {34},
pages = {2783-2791},
year = {2017},
issn = {0375-9601},
author = {Francisco Jauffred and Roberto Onofrio and Bala Sundaram}
}

@Article{Bohman2021,
author={Bohman, M.
and Grunhofer, V.
and Smorra, C.
and Wiesinger, M.
and Will, C.
and Borchert, M. J.
and Devlin, J. A.
and Erlewein, S.
and Fleck, M.
and Gavranovic, S.
and Harrington, J.
and Latacz, B.
and Mooser, A.
and Popper, D.
and Wursten, E.
and Blaum, K.
and Matsuda, Y.
and Ospelkaus, C.
and Quint, W.
and Walz, J.
and Ulmer, S.
and Collaboration, B. A. S. E.},
title={Sympathetic cooling of a trapped proton mediated by an LC circuit},
journal={Nature},
year={2021},
month={8},
day={01},
volume={596},
number={7873},
pages={514-518}
}

@article{Guggemos2015,
doi = {10.1088/1367-2630/17/10/103001},
url = {https://dx.doi.org/10.1088/1367-2630/17/10/103001},
year = {2015},
month = {9},
publisher = {IOP Publishing},
volume = {17},
number = {10},
pages = {103001},
author = {M Guggemos and D Heinrich and O A Herrera-Sancho and R Blatt and C F Roos},
title = {Sympathetic cooling and detection of a hot trapped ion by a cold one},
journal = {New Journal of Physics}
}

@article{Liu2023,
title = {Quantitative assessment and suppression of anharmonic potential of quadrupole linear radiofrequency ion traps with round electrodes},
journal = {International Journal of Mass Spectrometry},
volume = {485},
pages = {116997},
year = {2023},
issn = {1387-3806},
doi = {https://doi.org/10.1016/j.ijms.2022.116997},
url = {https://www.sciencedirect.com/science/article/pii/S1387380622002020},
author = {Y.H. Liu and L.J. Du and S.Y. Huang and Y.L. He and K.L. He and Q. Zhang and Y.L. Tang and Y.S. Meng and S.H. Zhai and H. Han and J. Xie}
}

@article{Sutherland2022,
  title = {One- and two-qubit gate infidelities due to motional errors in trapped ions and electrons},
  author = {Sutherland, R. Tyler and Yu, Qian and Beck, Kristin M. and H\"affner, Hartmut},
  journal = {Phys. Rev. A},
  volume = {105},
  issue = {2},
  pages = {022437},
  numpages = {13},
  year = {2022},
  month = {Feb},
  publisher = {American Physical Society}
}

@article{Lysne2024,
  title = {Individual Addressing and State Readout of Trapped Ions Utilizing Radio-Frequency Micromotion},
  author = {Lysne, Nathan K. and Niedermeyer, Justin F. and Wilson, Andrew C. and Slichter, Daniel H. and Leibfried, Dietrich},
  journal = {Phys. Rev. Lett.},
  volume = {133},
  issue = {3},
  pages = {033201},
  numpages = {6},
  year = {2024},
  month = {Jul},
  publisher = {American Physical Society}
}

@article{Gottesman2001,
  title = {Encoding a qubit in an oscillator},
  author = {Gottesman, Daniel and Kitaev, Alexei and Preskill, John},
  journal = {Phys. Rev. A},
  volume = {64},
  issue = {1},
  pages = {012310},
  numpages = {21},
  year = {2001},
  month = {Jun},
  publisher = {American Physical Society}
}

@Article{Flühmann2019,
author={Fl{\"u}hmann, C.
and Nguyen, T. L.
and Marinelli, M.
and Negnevitsky, V.
and Mehta, K.
and Home, J. P.},
title={Encoding a qubit in a trapped-ion mechanical oscillator},
journal={Nature},
year={2019},
month={Feb},
day={01},
volume={566},
number={7745},
pages={513-517}
}

@misc{Leon2024,
      title={Bosonic Quantum Error Correction with Neutral Atoms in Optical Dipole Traps}, 
      author={Leon H. Bohnmann and David F. Locher and Johannes Zeiher and Markus Müller},
      year={2024},
      eprint={2408.14251},
      archivePrefix={arXiv},
      primaryClass={quant-ph}
}

@article{Didier2012,
    author = {Didier, J. and Sinou, J.-J. and Faverjon, B.},
    title = {Multi-Dimensional Harmonic Balance With Uncertainties Applied to Rotor Dynamics},
    journal = {Journal of Vibration and Acoustics},
    volume = {134},
    number = {6},
    pages = {061003},
    year = {2012},
    month = {09},
    issn = {1048-9002}
}

@article{Chao2023,
title = {A state-of-the-art review on uncertainty analysis of rotor systems},
journal = {Mechanical Systems and Signal Processing},
volume = {183},
pages = {109619},
year = {2023},
issn = {0888-3270},
author = {Chao Fu and Jean-Jacques Sinou and Weidong Zhu and Kuan Lu and Yongfeng Yang}
}

@article{Miguel2022,
  title={Bayesian model identification through harmonic balance method for hysteresis prediction in bolted joints},
  author={Miguel, Luccas Pereira and Teloli, Rafael de Oliveira and Silva, Samuel da},
  journal={Nonlinear Dynamics},
  volume={107},
  number={1},
  pages={77--98},
  year={2022},
  publisher={Springer}
}

@article{sterk2022,
  title={Closed-loop optimization of fast trapped-ion shuttling with sub-quanta excitation},
  author={Sterk, Jonathan D and Coakley, Henry and Goldberg, Joshua and Hietala, Vincent and Lechtenberg, Jason and McGuinness, Hayden and McMurtrey, Daniel and Parazzoli, L Paul and Van Der Wall, Jay and Stick, Daniel},
  journal={npj Quantum Information},
  volume={8},
  number={1},
  pages={68},
  year={2022},
  publisher={Nature Publishing Group UK London}
}

@article{delaney2024,
  title={Scalable Multispecies Ion Transport in a Grid-Based Surface-Electrode Trap},
  author={Delaney, Robert D and Sletten, Lucas R and Cich, Matthew J and Estey, Brian and Fabrikant, Maya I and Hayes, David and Hoffman, Ian M and Hostetter, James and Langer, Christopher and Moses, Steven A and others},
  journal={Physical Review X},
  volume={14},
  number={4},
  pages={041028},
  year={2024},
  publisher={APS}
}

@article{blakestad2009,
  title={High-fidelity transport of trapped-ion qubits through an X-junction trap array},
  author={Blakestad, RB and Ospelkaus, C and VanDevender, AP and Amini, JM and Britton, Joseph and Leibfried, Dietrich and Wineland, David J},
  journal={Physical review letters},
  volume={102},
  number={15},
  pages={153002},
  year={2009},
  publisher={APS}
}

@misc{gesla2024,
      title={Stability analysis of periodic orbits in nonlinear dynamical systems using Chebyshev polynomials}, 
      author={Artur Gesla and Yohann Duguet and Patrick Le Quéré and Laurent Martin Witkowski},
      year={2024},
      eprint={2407.18230},
      archivePrefix={arXiv},
      primaryClass={physics.flu-dyn} 
}

@article{Xu2025,
doi = {10.1088/1361-6633/adb072},
url = {https://dx.doi.org/10.1088/1361-6633/adb072},
year = {2025},
month = {2},
publisher = {IOP Publishing},
volume = {88},
number = {3},
pages = {037602},
author = {Xu, Yingdan and Guo, Lingzhen},
title = {Perturbative framework for engineering arbitrary Floquet Hamiltonian},
journal = {Reports on Progress in Physics},
}

@article{Lignier2007,
  title = {Dynamical Control of Matter-Wave Tunneling in Periodic Potentials},
  author = {Lignier, H. and Sias, C. and Ciampini, D. and Singh, Y. and Zenesini, A. and Morsch, O. and Arimondo, E.},
  journal = {Phys. Rev. Lett.},
  volume = {99},
  issue = {22},
  pages = {220403},
  numpages = {4},
  year = {2007},
  month = {11},
  publisher = {American Physical Society}
}

@article{Schmied2009,
  title = {Optimal Surface-Electrode Trap Lattices for Quantum Simulation with Trapped Ions},
  author = {Schmied, Roman and Wesenberg, Janus H. and Leibfried, Dietrich},
  journal = {Phys. Rev. Lett.},
  volume = {102},
  issue = {23},
  pages = {233002},
  numpages = {4},
  year = {2009},
  month = {6},
  publisher = {American Physical Society}
}

@article{Li2012,
  title = {Space-Time Crystals of Trapped Ions},
  author = {Li, Tongcang and Gong, Zhe-Xuan and Yin, Zhang-Qi and Quan, H. T. and Yin, Xiaobo and Zhang, Peng and Duan, L.-M. and Zhang, Xiang},
  journal = {Phys. Rev. Lett.},
  volume = {109},
  issue = {16},
  pages = {163001},
  numpages = {5},
  year = {2012},
  month = {10},
  publisher = {American Physical Society}
}

@article{Qian2022,
  title = {Feasibility study of quantum computing using trapped electrons},
  author = {Yu, Qian and Alonso, Alberto M. and Caminiti, Jackie and Beck, Kristin M. and Sutherland, R. Tyler and Leibfried, Dietrich and Rodriguez, Kayla J. and Dhital, Madhav and Hemmerling, Boerge and H\"affner, Hartmut},
  journal = {Phys. Rev. A},
  volume = {105},
  issue = {2},
  pages = {022420},
  numpages = {10},
  year = {2022},
  month = {Feb},
  publisher = {American Physical Society}
}

@article{Paul1990,
  title = {Electromagnetic traps for charged and neutral particles},
  author = {Paul, Wolfgang},
  journal = {Rev. Mod. Phys.},
  volume = {62},
  issue = {3},
  pages = {531--540},
  numpages = {0},
  year = {1990},
  month = {7},
  publisher = {American Physical Society}
}

@article{kielpinski2002,
  title={Architecture for a large-scale ion-trap quantum computer},
  author={Kielpinski, David and Monroe, Chris and Wineland, David J},
  journal={Nature},
  volume={417},
  number={6890},
  pages={709--711},
  year={2002},
  publisher={Nature Publishing Group UK London}
}

@article{HAFFNER2008,
title = {Quantum computing with trapped ions},
journal = {Physics Reports},
volume = {469},
number = {4},
pages = {155-203},
year = {2008},
issn = {0370-1573},
author = {H. Häffner and C.F. Roos and R. Blatt},
keywords = {Quantum computing and information, Entanglement, Ion traps}
}

@article{lindblad2022,
  title={Minimizing aliasing in multiple frequency harmonic balance computations},
  author={Lindblad, Daniel and Frey, Christian and Junge, Laura and Ashcroft, Graham and Andersson, Niklas},
  journal={Journal of Scientific Computing},
  volume={91},
  number={2},
  pages={65},
  year={2022},
  publisher={Springer}
}

@article{Wang2012,
    author = {Wang, Xin and Li, Chuandong and Huang, Tingwen and Duan, Shukai},
    title = {Predicting chaos in memristive oscillator via harmonic balance method},
    journal = {Chaos: An Interdisciplinary Journal of Nonlinear Science},
    volume = {22},
    number = {4},
    pages = {043119},
    year = {2012},
    month = {11}
}

\title{
Supplemental Material for "Semi-Analytical Engineering of Strongly Driven Nonlinear Systems Beyond Floquet and Perturbation Theory"
}
\maketitle
\appendix
\onecolumngrid
\beginsupplement
\section{Example: Driven Optical Lattices}
Like Paul traps, anharmonicity modulation is valuable in optical lattice systems for various applications. Anharmonicity in optical lattice potentials can disrupt frequency-sensitive processes of the resolved-sideband cooling by detuning particle frequencies from laser resonance during cooling \cite{Siegel2024}. Conversely, anharmonicities play a crucial role in quantum simulations of p-band and d-band bosons in optical lattices by introducing the controllability on dephasing and degeneracy of the system \cite{Collin2010, Pinheiro2012, Pietraszewicz2013}. Therefore, anharmonicity modulation is crucial for efficiently cooling trapped atoms and realizing versatile bosonic quantum simulations in optical lattices.

Here, we propose a novel method to engineer individual standing wave potentials in optical lattices through lattice shaking. Remarkably, this approach can tune anharmonicities without requiring additional laser superposition, significantly simplifying experimental setups. Moreover, this method can extend to general optical dipole trap systems, such as optical tweezers, facilitating the creation of nonclassical motional states \cite{Leon2024}. 

We consider the one-dimensional Hamiltonian $(d = 1)$ for this driven optical lattice system given by:
\begin{equation}
    H = \frac{p^2}{2m} - \frac{V}{2} \cos{\left[2 k_L \left\{z - \Delta L \cos{\left(\Omega t \right)} \right\} \right]},
\end{equation}
where $p$, $m$, and $z$ are the momentum, the mass, and the position of a trapped neutral particle, respectively. Here, $V$ is the potential depth of an optical lattice, $k_L$ is the wave number of the counter-propagating laser light, and $\Delta L$ and $\Omega/2\pi$ are the amplitude and the frequency of the lattice shaking, respectively. Using the Hamilton equations $\dot{z} = \partial H/ \partial p$ and $\dot{p} = - \partial H/\partial z$, the equation of motion for the driven optical lattice becomes:
\begin{equation}
    \frac{d ^2 u}{d \xi^2} + V_0 \sin{\left[ 2 \left\{u - \lambda \cos{\left( 2 \xi \right)} \right\} \right]} = 0.
    \label{eomlattice}
\end{equation}
Here, $\xi = \Omega t/2$ is the normalized time, $u = k_L z$ is the dimensionless position, $\lambda = k_L \Delta L$ is the dimensionless drive amplitude, and $V_0 = 4 V k_L^2 / m \Omega^2$ is the effective potential depth.

\begin{figure}[t]
\includegraphics[width=8.4cm]{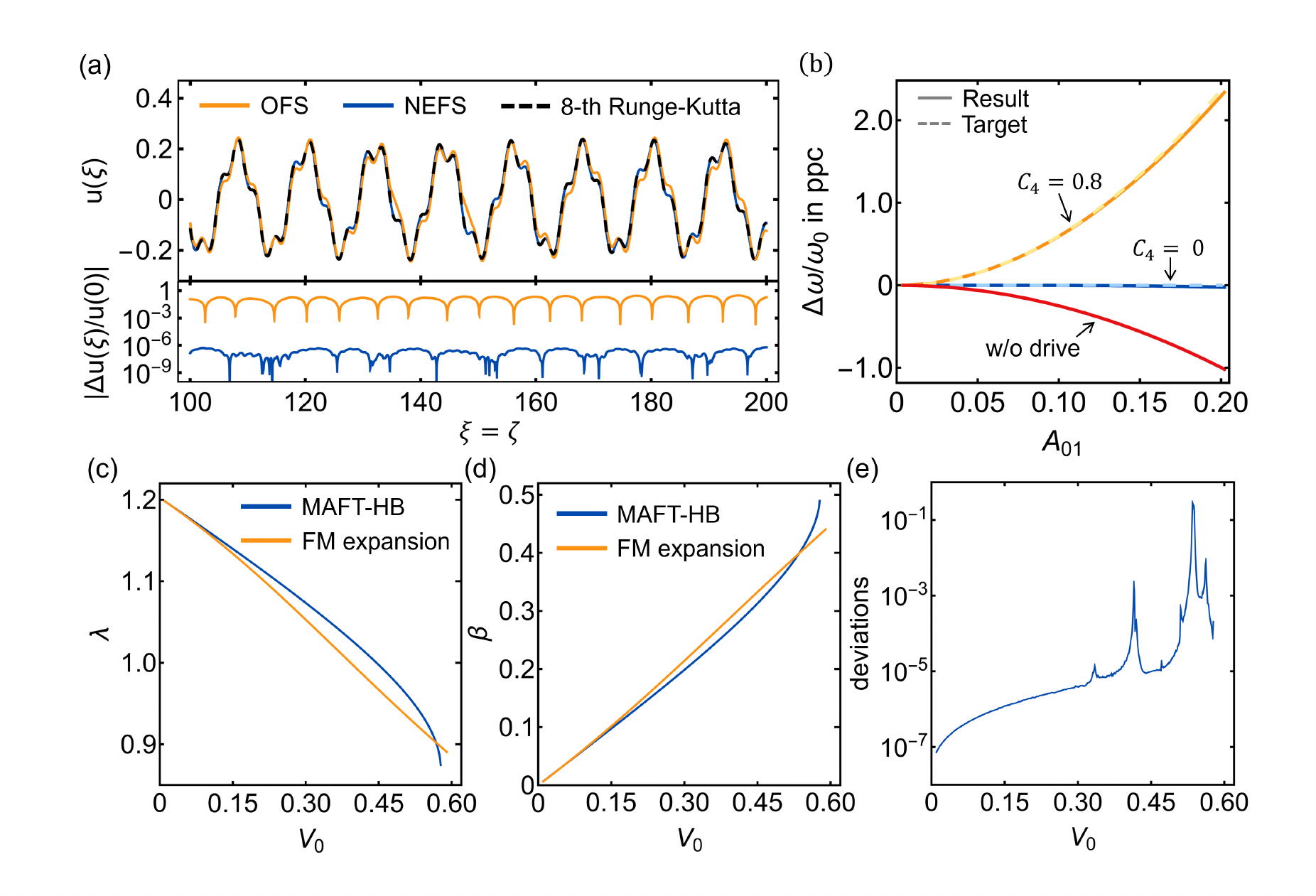}
\caption{\label{Fig2} Demonstration of semi-analytical engineering in driven optical lattices. (a) The trajectory of a trapped particle using the $8$-th Runge-Kutta method (dashed black) compared with results of the MAFT-HB method using OFS (orange) and NEFS (blue), along with their deviations (lower panel). (b) Relative amplitude-dependent frequency shifts of the target effective potentials (dashed lines), the engineered potentials (solid lines), and the shifts without drive (red). A ppc is $10^{-2}$. (c-e) Optimized drive amplitude $\lambda$ to achieve $C_4 = 0.8$, corresponding normalized secular frequency $\beta = 2\omega/\Omega$, and the maximum deviations from the numerical integration result as functions of the potential depth $V_0$.}
\end{figure}

We solved Eq.~(\ref{eomlattice}) with the MAFT-HB method to test the validity of NEFS, assuming $V_0 = 0.2$, $\lambda = 0.53$, and initial conditions $A_{01} = 0.2$, $\theta = 0$, where quasi-periodic oscillations and nonlinear effects dominate the trajectory. Here, we included $\sum_{m=-M}^{M} \tilde{A}_{m0} e^{-im\Omega \zeta} + \mathrm{c.c}$ with $\tilde{A}_{m0} = \frac{1}{2} A_{m0} e^{i \theta}, \ A_{m0} \in \mathbb{R}^d$ to NEFS and OFS, to incorporate the effect of the lattice shaking. Results were compared to those from the $8$-th order Runge-Kutta method in Figure~\ref{Fig2}(a). The NEFS matched the numerical integration more closely than the OFS, confirming the effectiveness of the nonlinear extended Floquet theory for the non-polynomial equation of motion. In the inverse problem, the driving amplitude $\lambda$ serves as the control parameter for potential engineering and was optimized to achieve target anharmonicities of $C_4 = 0.8, C_6 = C_8 = 0$ (orange), $C_4 = C_6 = C_8 = 0$ (blue), which result in the relative amplitude-dependent frequency shifts $\Delta \omega / \omega_0 = [\omega(A_{01}, \vb*{\varepsilon}) - \omega_0]/\omega_0$ of the dashed lines in Figure~\ref{Fig2}(b). Here, we solved $T(\vb*{a}_0, \vb*{a}_1; \vb*{b}_0, \vb*{b}_1) = O$ for $V_0 = 0.2$, $A_{01}^{(0)} = 10^{-5}$, and $ A_{01}^{(1)} = 10^{-4}$, and compared the relative amplitude-dependent frequency shifts of the optimized effective potentials (solid line) to targets and the shifts without drive ($\lambda = 0$, red line). The results matched the targets well but showed slight deviations at large amplitudes due to residual effects from higher-order anharmonicities  $(C_6, C_8, \dots)$, which cannot be fully controlled with a single control parameter. However, at small amplitudes, higher-order effects are negligible, and the accuracy of engineering $C_4$ was confirmed to be dominated by $|\Delta u (\xi)/u(0)|$, confirming the successful semi-analytical engineering of the effective potential. As demonstrated in Paul traps, introducing additional control parameters can further enhance the controllability of higher-order anharmonicities.

We compared the optimized driving amplitude $\lambda$ and the corresponding normalized secular frequency $\beta = 2 \omega/\Omega$ with those predicted by the second-order truncated FM expansion of Eq.~(\ref{Magnus}) for $0<V_0<0.6$, as shown in Figure~\ref{Fig2}(c,d). Significant discrepancies were observed at larger $V_0$, where weak and fast drive conditions break down. To ensure the accuracy of our method, we analyzed the maximum deviations  $|\Delta u (\xi)/u(0)|$ over $0 < \xi=\zeta < 200$, presented in Figure~\ref{Fig2}(e). While some deviation increase was observed at specific $V_0$ values due to higher-order harmonics not included in Eq.~(\ref{testfunction}), our method maintained high accuracy across the entire range of potential depths. These results demonstrate our method's robustness and universal applicability beyond the limitations of perturbative approaches.


\end{document}